\documentclass[showpacs,aps,prd,nofootinbib,floatfix,amsmath,amssymb]{revtex4}
\usepackage{graphicx}
\begin{document}

\makeatletter
\newbox\slashbox \setbox\slashbox=\hbox{$/$}
\newbox\Slashbox \setbox\Slashbox=\hbox{\large$/$}
\def\pFMslash#1{\setbox\@tempboxa=\hbox{$#1$}
  \@tempdima=0.5\wd\slashbox \advance\@tempdima 0.5\wd\@tempboxa
  \copy\slashbox \kern-\@tempdima \box\@tempboxa}
\def\pFMSlash#1{\setbox\@tempboxa=\hbox{$#1$}
  \@tempdima=0.5\wd\Slashbox \advance\@tempdima 0.5\wd\@tempboxa
  \copy\Slashbox \kern-\@tempdima \box\@tempboxa}
\def\FMslash{\protect\pFMslash}
\def\FMSlash{\protect\pFMSlash}
\def\miss#1{\ifmmode{/\mkern-11mu #1}\else{${/\mkern-11mu #1}$}\fi}
\makeatother

\title{Gauge structure of Yang-Mills theories with extra dimensions}
\author{J. A. Ahuatzi-Avenda\~no, A. J. Correa-Le\' on, M. Huerta-Leal, H. Novales-S\' anchez, A. Sierra-Mart\'\i nez, and J. J. Toscano}
\address{Facultad de Ciencias F\'{\i}sico Matem\'aticas,
Benem\'erita Universidad Aut\'onoma de Puebla, Apartado Postal
1152, Puebla, Puebla, M\'exico. }
\begin{abstract}
An effective Lagrangian for Yang-Mills theories with an arbitrary number of extra dimensions is constructed. We start from a field theory governed by the extra-dimensional Poincar\' e group ${\rm ISO}(1,3+n)$ and by the extended gauge group ${\rm SU}(N,{\cal M}^{4+n})$, which is characterized by an unknown energy scale $\Lambda$ and is assumed to be valid at energies far below this scale. Assuming that the size of the extra dimensions is much larger than the distance scale at which this theory is valid, an effective theory with symmetry groups ${\rm ISO}(1,3)$ and ${\rm SU}(N,{\cal M}^{4})$ is constructed. The transition between such theories is carried out via a canonical transformation that allows us to hide the extended symmetries ${\rm ISO}(1,3+n)\otimes {\rm SU}(N,{\cal M}^{4+n})$ into the standard symmetries ${\rm ISO}(1,3)\otimes {\rm SU}(N,{\cal M}^{4})$, and thus endow the Kaluza-Klein gauge fields with mass. Using a set of orthogonal functions $\{f^{(\underline{0})},f^{(\underline{m})}(\bar x)\}$, which is generated by the Casimir invariant $\bar {P}^2$ associated with the translations subgroup $T(n)\subset {\rm ISO}(n)$, the degrees of freedom of ${\rm ISO}(1,3+n)\otimes {\rm SU}(N,{\cal M}^{4+n})$ are expanded via a general Fourier series, whose coefficients are the degrees of freedom of ${\rm  ISO}(1,3)\otimes {\rm SU}(N,{\cal M}^{4})$. It is shown that these functions, which correspond to the projection on the coordinates basis $\{|\bar{x} \big >\}$ of the discrete basis $\{|0\big >,|p^{(\underline{m})}\big >\}$ generated by $\bar {P}^2$, play a central role in defining the effective theory. It is shown that those components along the base state $f^{(\underline{0})}=\big <\bar x|0\big>$ do not receive mass at the compactification scale, so they are identified with the standard Yang-Mills fields; but components along excited states $f^{(\underline{m})}=\big <\bar x|p^{(\underline{m})}\big>$ do receive mass at this scale, so they correspond to Kaluza-Klein excitations. In particular, it is shown that associated with any direction $|p^{(\underline{m})}\neq0\big >$ there are a massive gauge field and a pseudo-Goldstone boson. Some resemblances of this mass-generating mechanism with the Brout-Englert-Higgs mechanism are stressed.
\end{abstract}

\maketitle

\section{Introduction}
\label{I}The realization that there could be more than four spacetime dimensions goes back a long time ago when G. Nordstr\"om and T. Kaluza attempted to unify electromagnetism and gravity by assuming the existence of an extra spatial dimension~\cite{Nordstrom,Kaluza}. Nevertheless, it was O. Klein who found out, for the first time, that compactification could be used to explain the lack of observations of extra dimensions~\cite{Klein}. The ulterior birth of string theory, as a description of strong interactions~\cite{Veneziano,Nielsen,KoNi1,KoNi2,Nambu,Susskind1,Susskind2}, would eventually endow great relevance to formulations of extra dimensions. The original string-theory formulation already had this ingredient, as 26 spacetime dimensions were required to ensure unitarity~\cite{Lovelace}. The introduction of fermions in string theory~\cite{Ramond}, which came along with the discovery of  supersymmetry~\cite{Ramond,WeZu}, and the presence of a massless particle of spin 2~\cite{SchSch}, to be identified as the graviton, were two main elements of superstring theory that motivated its use to achieve a quantum theory of gravity, always with the complicity of extra dimensions. Remarkably, the critical dimension of superstring theory turned out to be just 10, as it was shown by J. H. Schwarz~\cite{Schwarz}. The introduction of the Green-Schwarz mechanism~\cite{GS}, to eliminate quantum anomalies arising in string theory, then triggered the first superstring revolution, during which five consistent superstring formulations were given~\cite{GS1,GHMR1,GHMR2,GHMR3}. Furthermore, a connection, through compactification, between superstring theory, featuring a 6-dimensional Calabi-Yau extra-dimensional manifold~\cite{Yau}, and 4-dimensional supersymmetry was established~\cite{CHSW}. A second superstring revolution started with the emergence of the $M$-theory, by E. Witten~\cite{Witten}, who showed that the five superstring  formulations known at the time are limits of this single theory, which is a unifying fundamental theory set in 11 spacetime dimensions. The existence of $D$-branes, proposed by J. Polchinski~\cite{Polchinski} for the sake of string duality, was a major event. It was also shown that supergravity in 11 dimensions is a low-energy limit of the $M$-theory~\cite{HoWi1,HoWi2}. The ADS/CFT correspondence, which establishes a duality of 5-dimensional theories of gravity with gauge field theories set in 4 dimensions~\cite{Maldacena}, is a quite important result with remarkable practical advantages regarding nonperturbative physics. Among the events and advances experienced by string theory throughout the years, and a plethora of papers on the matter, we wish to emphasize that its development is the one that got modern physics used to extra dimensions.
\\


The most emblematic feature of extra-dimensional field theories is, perhaps, the Kaluza-Klein (KK) towers, which involve, for each extra-dimensional field, an infinite set of 4-dimensional fields, the so-called KK modes, characterized by quantized mass spectra, determined by the compactification scale. The interest in phenomenological effects produced by new physics involving extra dimensions arose after the works by Antoniadis, Arkani-Hamed, Dimopoulos and Dvali~\cite{A,ADD,AADD}, who argued that relatively large extra dimensions could be detected at the TeV scale. In this new-physics approach, gravity is assumed to be the only interaction which propagates throughout the extra dimensions, thus explaining the relative weakness of gravitational interaction with respect to all other known interactions. So, the SM field content is assumed to lie exclusively in the ordinary 4-dimensional spacetime. In this framework, if only 1 extra dimension exists, its size would be $10^{\rm 13}\,{\rm cm}$~\cite{ADD}, which has been discarded, since this would imply deviations from Newtonian gravity at distances of order of the solar system. For the case of 2 extra dimensions their size is estimated to be $\sim10^{-2}\,{\rm cm}$~\cite{ADD}, then introducing deviations from Newton gravity at submillimeter distances. If more extra dimensions are assumed, they are predicted to be smaller. Deviations from Newton's gravitation law, generated by graviton KK modes, have been investigated, for instance, in Ref~\cite{torsionpendulum}, whose authors executed a torsion-pendulum experiment, thus arriving at the conclusion that the size of the extra-dimensional space, $R$, is restricted as $R<47\mu {\rm m}$ at $95\,\%$ CL, in the case of 2 equally-sized extra dimensions. An study, Ref.~\cite{CMSonADD}, by the ATLAS Collaboration has searched for the possibility of new physics lying in final states produced by proton-proton collisions, colliding at a center-of-mass energy of $\sqrt{s}=13\,{\rm TeV}$. By assuming a KK graviton to be in such final state, the interpretation of the results of this study has yielded the upper constraint $R<3.8\,\mu{\rm m}$, at $95\%$ CL~\cite{PDG}, again under the assumption that there are 2 extra dimensions. 
\\

Another approach, in which extra dimensions are assumed to be TeV-sized, was given by Randall and Sundrum, in what has been called ``models of warped extra dimensions''~\cite{RaSu1,RaSu2}. In the original version of this model, only gravity is assumed to propagate in the extra dimension of some 5-dimensional theory. This extra dimension is given an Anti-de-Sitter structure, which aims at explaining the conspicuously large difference among the electroweak and the Planck scales. Subsequent theoretical formulations of warped extra dimensions allowed for the propagation of SM fields in the whole 5-dimensional spacetime~\cite{DHR,Pomarol,CHNOY,GhPo}. A notorious alternative interpretation of warped extra-dimensions models links strongly-correlated theories, set in 4 dimensions, to 5-dimensional formulations with a warped extra dimension~\cite{Maldacena}. Under the assumption that only gravity propagates throughout the one warped extra dimension, the CMS Collaboration has performed an analysis of data from high-mass diphoton events, then arriving at a constraint as stringent as $4.6\,{\rm TeV}<m _{\rm KKG}^{(1)}$~\cite{CMSonRSgraviton}, on the mass of the lightest KK mode of the graviton, $m_{\rm KKG}^{(1)}$. Another investigation~\cite{CMSonRSk}, also performed by the CMS Collaboration, considers a version of the Randall-Sundrum model in which not only gravity but the whole SM can propagate in the extra dimension. This study analyses data from dilepton final states, originated in proton-proton collisions, and then finds constraints on the warp parameter, $k_{\rm RS}$, which turn out to be as restrictive as $4.78\leqslant\frac{k_{\rm RS}}{\overline{M}_{\rm Pl}}$, with $\overline{M}_{\rm Pl}$ the reduced Planck mass. The mass of the first KK mode of the gluon has also been constrained in an investigation carried out by the ATLAS Collaboration~\cite{ATLASonRSgluon}. By assuming that top-quark pairs are produced by the decays of new particles, the ATLAS Collaboration finds that the mass of the first KK mode of the gluon is lower-bounded as $m_{g}^{(1)}>3.4\,{\rm TeV}$. 
\\

A further relevant model of extra dimensions is the one dubbed ``universal extra dimensions''~\cite{ACD1}, in which the extra dimensions are assumed to be spacelike and flat, with all the SM field content allowed to propagate in the whole spacetime. One of the main motivations behind the approach of universal extra dimensions is that it provides a viable dark-matter candidate~\cite{ChMS,CFM,SeTa1,SeTa2,HooKr,BHS,KoMa,HooPr,DHKM,BNP,BKP,BMMO}, a role played by the lightest KK mode of some neutral field, which could be, for instance, the photon or some neutrino. Notably, models of universal extra dimensions introduce only one new parameter, namely, the compactification scale, $R^{-1}$, which makes them highly predictive. Ref.~\cite{DFK} considers supersymmetry searches by the LHC at center-of-mass energies of $8\,{\rm TeV}$ and $13\,{\rm TeV}$, which is then used to set the lower bound $1.4\,{\rm TeV}\lesssim R^{-1}$. Besides this, that paper establishes that consistency of this model with cosmology restricts the compactification scale to be within $1.25\,{\rm TeV}\lesssim R^{-1}\lesssim1.5\,{\rm TeV}$. Similar results have been reached by the authors of Ref.~\cite{BDDM}. An investigation, by the authors of Ref.~\cite{FKRR}, centered on the analysis of data from the Large Hadron Collider, considering the Run 2. The authors of that work conclude that lower bounds as restrictive as $1.8\,{\rm TeV}\lesssim R^{-1}$, on the compactification scale $R^{-1}$, can be inferred. They also claim that this bound rules out the minimal model of 1 universal extra dimension if standard cosmology is assumed to be correct. 
\\

If large extra dimensions exist, they can affect low energy processes. Therefore, it is interesting to construct an effective theory that allows us to explore their impact on low-energy observables. The main goal of this paper is to discuss the theoretical bases that lead to the formulation of such an effective theory. The aspects which bear the most interest, but at the same time the ones characterized by the greatest intricacies, have to do with gauge invariance, so we will focus our discussion on a pure (without matter fields) Yang-Mills theory. Our study will be based on the well-known universal extra dimensions approach~\cite{ACD1}. A study at the quantum level of the material presented here has been already reported by some of us~\cite{PRDYM}. Also, other results that are closely related with this work have been discussed by some of us in Refs.~\cite{PRDQED,JPGHIGGS,JPGS,OPT1,OPT2,OPT3,FGNT,MNT,GNT}. However, these works do not exhaustively address the structural bases of the theory at the classical level. Our purpose is to deepen into these ideas and, at the same time, to present a unified version of this topic at the classical level. Firstly, we will study the structure of the effective theory that results from the compactification and integration of the extra dimensions, identifying their degrees of freedom, gauge invariances and mass spectrum. \\

Our approach will be based on the notion of ``hidden symmetry'', a concept which serves as a powerful instrument that allows to elegantly describe some subtle and complex scenarios of fundamental physics. Hidden symmetry and spontaneous symmetry breaking (SSB)~\cite{SSB} are the cornerstone of the Standard Model (SM) of particle physics. The merger of both ideas is the very essence of the famous Brout-Englert-Higgs mechanism (BEHM)~\cite{EHM1,EHM2,EHM3}. In this work, we follow this approach closely to lay the foundations of a Yang-Mills extension to extra flat dimensions. We consider, as a starting point, an action for a field theory defined on a flat-spacetime manifold: ${\mathcal M}^{d}={\mathcal M}^4\otimes {\mathcal N}^n$, which is the result of the cartesian product of the four-dimensional Minkowski spacetime ${\cal M}^4$ and some $ n $-dimensional noncompact manifold ${\cal N}^n$ that represents a spatial extension. We assume this higher-dimensional theory to be invariant under the Poincar\'e group ${\rm ISO}(1,3+n)$ and the gauge group ${\rm SU}(N,{\mathcal M}^{d})$, which is a gauge group with all of its group parameters defined on $ \mathcal{M}^{d} $. Since this theory is not renormalizable in the Dyson's sense, it is given by a Lagrangian that includes an infinite number of ${\rm ISO}(1,3+n)\otimes {\rm SU}(N,{\cal M}^d)$-invariant terms of all possible canonical dimensions. The lowest-dimensional term corresponds to a direct $(4+n)$-dimensional extension of the standard $4$-dimensional theory, while those terms of increasing canonical dimension are suppressed by inverse powers of an unknown energy scale $\Lambda$, which is assumed to be far above the compactification scale $R^{-1}$. We assume that the average size of the extra dimensions, $R$, is so large compared with the distance scale at which this theory is valid that extra dimensions can be practically considered as infinite. Thus, at energies far above the compactification scale $R^{-1}$, this theory is governed by the extended groups ${\rm ISO}(1,3+n)\otimes {\rm SU}(N,{\cal M}^d)$. To describe the physical phenomena at much-smaller energies, where the compactness of the extra dimensions becomes apparent, we need to hide the ${\rm ISO}(1,3+n)\otimes {\rm SU}(N,{\cal M}^d)$ symmetry into ${\rm ISO}(1,3)\otimes {\rm SU}(N,{\cal M}^4)$. Observe that ${\rm SU}(N,{\cal M}^d)$ and ${\rm SU}(N,{\cal M}^4)$ coincide as Lie groups, but they differ as gauge groups. It should be noted that the process of hiding a symmetry does not mean moving from one theory to another, but rather focusing on the same theory from another perspective. This means that we must pass from the description based on ${\rm ISO}(1,3+n)\otimes {\rm SU}(N,{\cal M}^d)$ to that characterized by ${\rm ISO}(1,3)\otimes {\rm SU}(N,{\cal M}^4)$ through a canonical transformation. As it occurs in theories with SSB, the physical content is a matter of scales. In the SM one uses the groups ${\rm SU}(3,{\cal M}^4)_C\otimes {\rm U}(1,{\cal M}^4)_e$ to describe physical phenomena at energies of the order of the Fermi scale $v$, but at energies far above $v$ the ${\rm SU}(3,{\cal M}^4)_C\otimes {\rm SU}(2,{\cal M}^4)_L\otimes {\rm U}(1,{\cal M}^4)_Y$ description must be used. In our case, at energies of the order of the compactification scale $R^{-1}$, we use the ${\rm ISO}(1,3) \otimes {\rm SU}(N,{\cal M}^4)$ description. However, at energies far above the $R^{-1}$ scale, we use the ${\rm ISO}(1,3+n) \otimes {\rm SU}(N,{\cal M}^d)$ description, since at these energies we are exploring distances so small that the compact dimensions would really look infinite.
\\

The compactification program comprises a number of nontrivial steps. First, one must define a canonical transformation that maps covariant objets of $ {\rm ISO}(1,3+n) \otimes {\rm SU}(N,{\cal M}^d)$ into covariant objets of ${\rm ISO}(1,3) \otimes {\rm SU}(N,{\cal M}^4)$. This transformation is crucial to hide the extended symmetry into the standard one. Second, since the number of connections of ${\rm SU}(N,{\cal M}^4)$ is smaller than that of ${\rm SU}(N,{\cal M}^d)$, the difference will appear in tensorial representations of ${\rm SU}(N,{\cal M}^4)$. So from the perspective of the standard ${\rm SU}(N,{\cal M}^4)$ gauge group, these connections can be endowed with mass. Indeed, it is necessary to endow such connections with mass at the compactification scale $R^{-1}$ because these new-physics effects must be of decoupling nature in accordance with the Appelquist-Carazzone's decoupling theorem~\cite{AC}. This means that some instrument analogous to SSB must be available in order to generate such masses. Any new particle that emerges as a consequence of the process of hiding the symmetry must be endowed with mass through that instrument. The pass from the extended symmetry to the standard one should not spoil the gauge structure of the theory, which means that we must be able to examine the physical phenomenon from both the ${\rm ISO}(1,3) \otimes {\rm SU}(N,{\cal M}^4)$ (low-energies) and  ${\rm ISO}(1,3+n) \otimes {\rm SU}(N,{\cal M}^d)$ (high-energies) points of view. As it occurs in theories with SSB, the hiding of the gauge symmetry manifests itself through the presence of two types of gauge transformations: the standard gauge transformations (SGTs) of the group ${\rm SU}(N,{\cal M}^4)$ and a set of nonstandard gauge transformations (NSGTs). The NSGTs are determined by gauge parameters which do not belong to the ${\rm SU}(N,{\cal M}^4)$ gauge group (see Refs.~\cite{PRDYM,PRDQED,JPGHIGGS,OPT1,OPT2,OPT3}). A nontrivial consequence of hiding the extended symmetry into the standard one is the presence of an infinite number of basic fields produced by the canonical map. These fields correspond to the so-called KK towers; some of these towers involve a field that can be identified with a standard gauge field on $ \mathcal{M}^{4} $ because it does not receive mass from the compactification mechanism, while all the remaining KK fields receive mass through this mechanism.
\\

As we commented above, our approach is based on the notion of hidden symmetry. Since this concept is central to the SM, one of our main purposes is to establish a parallelism of the KK mass generating mechanism with the BEHM, highlighting both similarities and discrepancies. An important goal is to establish the sequence of hiding symmetries:
\begin{equation}
\label{BS}
{\rm ISO}(1,3+n)\otimes {\rm SU}(N,{\cal M}^d)  \xrightarrow{R^{-1}} {\rm ISO}(1,3)\otimes {\rm SU}(N,{\cal M}^4) \, .
\end{equation}
\\

The rest of the paper has been organized as follows. In Sec.~\ref{P}, a qualitative discussion about the BEHM at the classical level is presented. The implications of the Lie algebra of the ${\rm ISO}(1,3+n)$ group and its subgroups ${\rm ISO}(1,3)$ and ${\rm ISO}(n)$ on the constants of motion of the system are discussed. In Sec. \ref{GF}, the conceptual and technical ingredients behind the derivation of the extended four-dimensional Yang-Mills theory are presented. In Sec. \ref{C}, a summary of our results is presented. In Appendices~\ref{S} and~\ref{AE} we collect several conventions and integrals that have been used in our derivations. \\

\section{Preliminaries}
\label{P}One of the purposes of this section is to present a qualitative discussion, at the classical level, of the main features of the BEHM. Another point to be discussed has to do with the Lie algebra of the spacetime groups and their associated constants of motion.

\subsection{The Brout-Englert-Higgs Mechanism}
In order to establish an appropriate context for our discussion, let us briefly recall the main ingredients of theories with SSB. The degrees of freedom of some gauge field theory under consideration are characterized by connections and tensorial representations of a gauge group $G$. In the jargon of particle physics, the former are known as ``gauge fields'' and the latter are called ``matter fields''. This classification has profound consequences at the quantum level because gauge symmetry only allows mass terms for those fields that appear in tensorial representations of $G$. Mass terms for gauge fields spoil gauge invariance, so, from the $G$-group perspective, they are necessarily associated with massless particles. To endow with mass this class of fields, we need to approach the theory from another perspective, in which some of the gauge fields of $G$ look like matter fields. This change of perspective can be implemented through a canonical transformation that allows us to map covariant objets of $G$ into covariant objets of one of its subgroups, $H$. Since the number of connections of $H$ is smaller than that of $G$, the connections that do not belong to $H$ will appear in tensorial representations of it. So, to endow some of the connections of $G$ with mass, one needs to hide the $G$ symmetry into the $H$ symmetry. This does not mean that the $G$ symmetry is lost, since the canonical map can be reversed. The concept of hidden symmetry is central to the BEHM~\cite{EHM1,EHM2,EHM3}, which allows us to generate masses for gauge fields through the phenomenon of SSB~\cite{SSB}.\\

The SSB of a continuous symmetry occurs in the presence of an infinitely-degenerate vacuum, which has its origin in a scalar potential with very peculiar features. In the simplest scenario, this potential defines a spherical surface of radius $v$ on which the system has its minimal energy. All the points on the sphere are  physically equivalent because they are related by the $G$ group. To breakdown $G$ into $H$, a particular point of the sphere, characterized by a constant vector $\Phi_0$, must be chosen. It is said that the group $G$ is broken into its subgroup $H$ at the scale $v$, in the sense that the $\Phi_0$ direction is left invariant by the $H$ group, that is, $\Phi_0$ is annihilated by the generators $T^{\bar{a}}$ of $H$. If the symmetry is global, we have the Goldstone theorem, which tells us that for each broken generator of $G$ (those generators $T^{\hat{a}}$ of $G$ such that $T^{\hat{a}}\Phi_0\neq 0$) there is an associated massless scalar particle (Goldstone boson). If $G$ is a gauge group, the gauge fields associated  with the broken generators of $G$ acquire a mass proportional to the scale $v$. This is the celebrated BEHM. Its physical meaning is very interesting: at energy scales of order of $v$, the phenomenology is explained by the $H$ group, but at energies far above of $v$, we must use the $G$ perspective to describe the physical phenomena. Thus, whether we describe phenomena through the group $G$ or the group $H$ is a matter of energy scales. Below, we follow this approach closely to construct a Yang-Mills extension to extra dimensions.

\subsection{Spacetime constants of motion}

In a flat $d$-dimensional spacetime, with $d=4+n$, the Poincar\' e group ${\rm ISO}(1,d-1)$ is defined through its generators, whose number is equal to $\frac{1}{2}d(d+1)$. $d$ of such generators, denoted by $P_M$, belong to the group of translations, ${\rm T}(1,d-1)$, while the remaining $\frac{1}{2}d(d-1)$ generators, denoted by $J_{MN}$, are associated with the Lorentz group ${\rm SO}(1,d-1)$. These generators satisfy the Poincar\' e algebra
\begin{eqnarray}
\label{PL1}
&&[P_M\, ,\, P_N]=0\, , \\
\label{PL2}
&&[J_{MN}\, ,\, P_R]=i\left(g_{MR}P_{N}-g_{NR}P_M\right)\, , \\
\label{PL3}
&&[J_{MN}\, , \, J_{RS}]=i\left(g_{MR}J_{NS}-g_{MS}J_{NR}-g_{NR}J_{MS}+g_{NS}J_{MR}\right)\, .
\end{eqnarray}
It is not difficult to see that there are two merged subalgebras in this algebra. One of them corresponds to the standard Poincar\' e group ${\rm ISO}(1,3)$,  \begin{eqnarray}
&&[P_\mu \, ,\, P_\nu ]=0\, , \\
&&[J_{\mu \nu }\, ,\, P_\rho]=i\left(g_{\mu \rho}P_{\nu }-g_{\nu \rho}P_\mu \right)\, , \\
&&[J_{\mu \nu}\, , \, J_{\rho \sigma}]=i\left(g_{\mu \rho}J_{\nu \sigma}-g_{\mu \sigma}J_{\nu \rho}-g_{\nu \rho}J_{\mu \sigma}+g_{\nu \sigma}J_{\mu \rho}\right)\, ,
\end{eqnarray}
whereas the other is associated with the inhomogeneous orthogonal group in $n$ dimensions, ${\rm ISO}(n)$:
\begin{eqnarray}
\label{IN1}
&&[P_{\bar{\mu}}\, , \,P_{\bar{\nu}}]=0\, , \\
\label{IN2}
&&[J_{\bar{\mu}\bar{\nu}}\, , \,P_{\bar{\rho}}]=i\left(\delta_{\bar{\nu}\bar{\rho}}P_{\bar{\mu}}-\delta_{\bar{\mu}\bar{\rho}}P_{\bar{\nu}}\right)\, ,\\
&&[J_{\bar{\mu}\bar{\nu}}\, , \, J_{\bar{\rho}\bar{\sigma}}]=i\left( \delta_{\bar{\mu}\bar{\sigma}}J_{\bar{\nu}\bar{\rho}}-\delta_{\bar{\mu}\bar{\rho}}J_{\bar{\nu}\bar{\sigma}}-\delta_{\bar{\nu}\bar{\sigma}}J_{\bar{\mu}\bar{\rho}}+
\delta_{\bar{\nu}\bar{\rho}}J_{\bar{\mu}\bar{\sigma}}\right) \, .
\end{eqnarray}

By identifying $P_0$ with the Hamiltonian of the system, we can see, from Eqs.~(\ref{PL1}) and (\ref{PL2}), that the generators $P_\mu$, $P_{\bar \mu}$, $J_{ij}$, and $J_{\bar \mu \bar \nu}$ are all constants of motion. This in turn implies that all the generators of the inhomogeneous ${\rm ISO}(n)$ group are constants of motion. As we will see later, this fact plays a central role in the mechanism to generate the KK mass spectrum. On the other hand, it is easy to see that the generators of boosts $J_{0i}$ and $J_{0\bar \mu}$ are not conserved quantities. Below, we will profit from this discussion on constants of motion; the Casimir invariant $P_{\bar{\mu}}P_{\bar{\mu}}$ will be used to define a complete set of orthogonal functions that determines (1) a canonical map to change the perspective of the theory from extra dimensions to four dimensions, and (2) a mass spectrum for KK modes that is in accordance with the decoupling theorem.

\section{Yang-Mills theories with extra dimensions}
\label{GF}
The starting point is an effective gauge field theory governed by the extended group ${\rm ISO}(1,3+n) \otimes {\rm SU}(N,{\mathcal M}^{d})$, and whose gauge parameters are defined all over the spacetime $\mathcal{M}^{d}=\mathcal{M}^{4}\otimes\mathcal{N}^{n} $. The action of the theory is assumed to be a functional of gauge fields. Since the theory is not renormalizable in the Dyson's sense, the corresponding action consists of an infinite series of Lorentz and gauge invariant terms of increasing canonical dimension, that is,
\begin{equation}\label{ea}
S_{\rm eff}[{\cal A}^a_M]=\int d^{\,4}x\,d^{\,n}\bar{x} \,\, {\cal L}_{4+n}\left({\cal F}^a_{MN},{\cal D}^a_A {\cal F}^a_{MN},\cdots \right)\, ,
\end{equation}
where
\begin{eqnarray}
\label{LG}
 {\cal L}_{4+n}=-\frac{1}{4}{\cal F}^a_{MN}{\cal F}^{MN}_a+\sum_{\textbf{d}} \frac{\lambda_{\textbf{d}}}{\Lambda^{\textbf{d}}}\,{\cal L}^{(\textbf{d}+d)}\left({\cal F}^a_{MN},{\cal D}^a_A {\cal F}^a_{MN},\cdots \right)\, ,
\end{eqnarray}
with
\begin{equation}
\label{YMC}
{\cal F}^a_{MN}(x,\bar x)=\partial_M {\cal A}^a_N(x,\bar x)-\partial_N {\cal A}^a_M(x,\bar x)+g_{4+n}f^{abc}{\cal A}^b_M(x,\bar x){\cal A}^c_N(x,\bar x)\, .
\end{equation}
In this expression, $g_{4+n}$ and $f^{abc}$ are the coupling constant and the structure constant of the ${\rm SU}(N,{\cal M}^d)$ group, respectively. Under the extended gauge group ${\rm SU}(N,{\cal M}^d)$, the connection and curvature components transform as
\begin{subequations}
\begin{align}
\label{GT}
\delta {\cal A}^a_M(x,\bar x)&={\cal D}^{ab}_M \alpha^b(x,\bar x)\, , \\
\label{CT}
\delta {\cal F}^a_{MN}(x,\bar x)&=g_{4+n}f^{abc} {\cal F}^b_{MN}(x,\bar x)\alpha^c(x,\bar x)\, ,
\end{align}
\end{subequations}
where $\alpha^a(x,\bar x)$ are the gauge parameters and ${\cal D}^{ab}_M=\delta^{ab}\partial_M-g_{4+n}f^{abc}{\cal A}^c_M(x,\bar x)$ is the covariant derivative in the adjoint representation of the group. In Eq.~(\ref{LG}), ${\cal L}^{(\textbf{d}+d)}$ represents gauge- and Lorentz-invariant interactions of canonical dimension greater than $d$, formulated from curvatures ${\cal F}^a_{MN}$ and its covariant derivatives ${\cal D}^{ab_1}_{M_1}{\cal D}^{b_1b_2}_{M_2}\cdots{\cal D}^{b_{k-1}b_k}_{M_k} {\cal F}^{b_k}_{MN}$, multiplied by unknown coupling constants $ \lambda_{\textbf{d}}/\Lambda^{\textbf{d}} $. The first term in the right-hand side of Eq.~\eqref{LG} corresponds to a straightforward extension of the functional structure of the well-known Yang-Mills Lagrangian from four dimensions to $ d $ dimensions. Terms of greater canonical dimension are suppressed by inverse powers of a fundamental scale $\Lambda$, which is assumed to be far above the explored energies. In fact, according to the effective Lagrangian approach, this theory is valid only for energies $<\Lambda$. We assume that the size of the extra dimensions is so large compared with these energies that it actually can be considered as infinite. This justifies the ${\rm ISO}(1,3+n)$ description of the effective theory given by Eq.~\eqref{LG}. Note that the first term in Eq.~\eqref{LG} does not depend on the $\Lambda$ scale, although it does depend on the dimensionful coupling constant $g_{4+n}$, which must be rescaled to obtain the correct dimensionless Yang-Mills coupling. As we see below, this term plays a central role in the ${\rm ISO}(1,3) \otimes {\rm SU}(N,{\mathcal M}^{4})$ description.
\\

Our main goal is to construct a Yang-Mills extension defined in extra dimensions. To do this, we need to define canonical maps that allow us to descend towards low-energy regimes in accordance with the pattern of symmetries hiding that we schematized in Eq.~(\ref{BS}). To carry out this program, we stress that $ {\rm SO}(1,3) $ and $ {\rm SO}(n) $ are subgroups of $ {\rm SO}(1,3+n) $. Also, note that the gauge fields ${\cal A}^a_M(x,\bar x)$, which are vector fields of $ {\rm SO}(1,3+n) $, can be seen as an $ {\rm SO}(1,3) $-vector field, with components $ \mathcal{A}^a_{\mu} $, and  $ n $ $ {\rm SO}(1,3) $-scalars denoted by $ \mathcal{A}^a_{\bar{\mu}} $; under the ${\rm SO}(n)$ group, the four components of $ \mathcal{A}^a_{\mu} $ are scalar fields, whereas $ \mathcal{A}^a_{\bar{\mu}} $ can be seen as a vector.\\
\\

\subsection{Hiding the symmetry}
\label{hs}
 From now on, we will focus on the first term of the effective Lagrangian given by Eq.~(\ref{LG}), which defines the action
 \begin{equation}
 \label{A1}
 S[{\cal A}^a_M]=-\frac{1}{4}\int d^{\,4}x\,d^{\,n}\bar{x}\,{\cal F}^a_{MN}(x,\bar x){\cal F}^{MN}_a(x,\bar x) \, .
\end{equation}
To hide the symmetry ${\rm ISO}(1,3+n) \otimes {\rm SU}(N,{\cal M}^d)$ into ${\rm ISO}(1,3) \otimes {\rm SU}(N,{\cal M}^4)$ means to hide the ${\rm ISO}(1,3+n)$ symmetry into ${\rm ISO}(1,3)$, since implicit to this is the passing from ${\rm SU}(N,{\cal M}^d)$ to ${\rm SU}(N,{\cal M}^4)$. We first map covariant objets of $ {\rm SO}(1,3+n) $ into covariant objets of its subgroups ${\rm SO}(1,3)$ and ${\rm SO}(n)$:
\begin{eqnarray}
\label{MF1}
{\rm SO}(1,3+n) &\mapsto & {\rm SO}(1,3)\times {\rm SO}(n) \, ,\nonumber  \\
{\cal A}^a_M(x,\bar x)  &\mapsto & \{ {\cal A}^a_\mu (x,\bar x), \, {\cal A}^a_{\bar \mu} (x,\bar x)\}\, .
\end{eqnarray}
This map in turn leads to the curvatures
\begin{equation}
\label{MC1}
{\cal F}^a_{MN}(x,\bar x) \mapsto \{ {\cal F}^a_{\mu \nu}(x,\bar x), \, {\cal F}^a_{\mu \bar \nu }(x,\bar x), \, {\cal F}^a_{\bar \mu \bar \nu}(x,\bar x)\} \, ,
\end{equation}
 with ${\cal F}^a_{\mu \nu}(x,\bar x)$, ${\cal F}^a_{\mu \bar \nu }(x,\bar x)$, and ${\cal F}^a_{\bar \mu \bar \nu}(x,\bar x)$ transforming as 2-form (0-form), 1-form (1-form), and 0-form (2-form) under the ${\rm SO}(1,3)({\rm SO}(n))$ group, respectively.
 \\

The maps given by Eqs.~(\ref{MF1}) and (\ref{MC1}) allow us to pass from the action given in Eq.~(\ref{A1}) to an action given by
\begin{eqnarray}
\label{A2}
S[{\cal A}^a_\mu,{\cal A}^a_{\bar \mu}]&=&-\frac{1}{4} \int d^4x\, d^n \bar x \Big[{\cal F}^a_{\mu \nu}(x,\bar x){\cal F}^{\mu \nu}_a(x,\bar x)+2{\cal F}^a_{\mu \bar \nu}(x,\bar x){\cal F}^{\mu \bar \nu}_a(x,\bar x)\nonumber \\
&&+{\cal F}^a_{\bar \mu \bar \nu}(x,\bar x){\cal F}^{\bar \mu \bar \nu}_a(x,\bar x)
\Big]\, ,
\end{eqnarray}
which is manifestly invariant under the ${\rm SO}(1,3)$ and ${\rm SO}(n)$ groups. Although from the ${\rm SO}(1,3)$ perspective the extra-dimensional symmetry ${\rm SO}(1,3+n)$ seems to have disappeared, this is not actually the case, as one can trivially pass from the action (\ref{A2}) to the action (\ref{A1}), in which the larger symmetry is manifest. What really happened is that the map (\ref{MF1}) hid the ${\rm SO}(1,3+n)$ symmetry. We can see that this map splits Eq.(\ref{GT}) into two identical expressions for the ${\cal A}^a_\mu(x,\bar x)$ and ${\cal A}^a_{\bar \mu}(x,\bar x)$ connection components; whereas Eq.(\ref{CT}) has been split into three identical expressions for the curvature components $\{{\cal F}^a_{\mu \nu}(x,\bar x), \, {\cal F}^a_{\mu \bar \nu }(x,\bar x), \, {\cal F}^a_{\bar \mu \bar \nu}(x,\bar x)  \}$. This means that the action (\ref{A2}) remains governed by the gauge group ${\rm SU}(N,{\cal M}^d)$. Evidently, this point map can easily be elevated to a canonical map at the phase space.
\\

The map given in Eq.~(\ref{MF1}) accommodates representations of ${\rm SO}(1,3+n)$ into representations of the groups ${\rm SO}(1,3)$ and ${\rm SO}(n)$. However, to move completely from the extended symmetry to the standard one, we need to remove any dependence on the $\bar x$ coordinates from the theory. This step is nontrivial because, in the original theory, these coordinates play the role of labels that count degrees of freedom. In other words, we need to define another canonical map that allows us to hide any manifest dynamical role of the ${\rm ISO}(n)$ subgroup. To do this, assume that some compactification procedure on the ${\cal N}^n$ submanifold has been carried out, and let $\{ f^{(\underline{m})}(\bar x)\}$ be a complete set of orthogonal functions defined on the compact manifold. Then, the fields appearing in Eq.~(\ref{MF1}) and the gauge parameters, $\alpha^a(x,\bar x)$, can be decomposed in this basis as follows:
\begin{subequations}
\begin{align}
\label{MFV2}
{\cal A}^a_\mu(x,\bar x)&=\sum_{(\underline{m})}f^{(\underline{m})}(\bar x)A^{(\underline{m})a}_\mu(x)\, , \\
\label{MFS2}
{\cal A}^a_{\bar \mu}(x,\bar x)&=\sum_{(\underline{m})}f^{(\underline{m})}(\bar x)A^{(\underline{m})a}_{\bar \mu}(x)\, , \\
\label{MGP}
\alpha^a(x,\bar x)&=\sum_{(\underline{m})}f^{(\underline{m})}(\bar x)\alpha^{(\underline{m})a}(x)\, ,
\end{align}
\end{subequations}
with analogous expressions for $\Pi^a_\mu$ and $\Pi^a_{\bar \mu}$, which are the canonical conjugates of ${\cal A}^a_\mu(x,\bar x)$ and ${\cal A}^a_{\bar \mu}(x,\bar x)$, respectively. The symbol $\sum_{(\underline{m})}$ is defined in Appendix~\ref{S}. In the above map, the degrees of freedom are characterized by the infinite set of fields $\{A^{(\underline{m})a}_\mu(x), \, A^{(\underline{m})a}_{\bar \mu}(x) \}$, while the $f^{(\underline{m})}(\bar x)$ functions do not represent degrees of freedom. Assuming that the functions of the set $\{ f^{(\underline{m})}(\bar{x}) \}$ are normalized, the above maps can be reversed to obtain
\begin{subequations}
\begin{align}
\label{I1}
A^{(\underline{m})a}_\mu(x)&=\int d^n\bar x\, f^{(\underline{m})}(\bar x){\cal A}^a_\mu(x,\bar x)\, , \\
A^{(\underline{m})a}_{\bar \mu}(x)&=\int d^n\bar x\, f^{(\underline{m})}(\bar x){\cal A}^a_{\bar \mu}(x,\bar x)\, .
\end{align}
\end{subequations}
On the other hand, using the orthonormality of the $f^{(\underline{m})}(\bar x)$ functions, it is easy to show that the fundamental Poisson's brackets
\begin{subequations}
\begin{align}
\{{\cal A}^a_\mu(x,\bar x), \, \Pi^b_\nu(x',\bar x')\}&=\delta_{ab}\, \delta_{\mu \nu}\,\delta(\mathbf{x}-\mathbf{x'})\, \delta(\bar x-\bar x')\, ,\\
\{{\cal A}^a_{\bar \mu}(x,\bar x), \, \Pi^b_{\bar \nu}(x',\bar x')\}&=\delta_{ab}\, \delta_{\bar \mu \bar \nu}\,\delta(\mathbf{x}-\mathbf{x'})\, \delta(\bar x-\bar x')\, ,
\end{align}
\end{subequations}
become
\begin{subequations}
\begin{align}
\{A^{(\underline{m})a}_\mu(x), \, \pi^{(\underline{n})b}_\nu(x')\}&=\delta_{ab}\, \delta_{\mu \nu}\, \delta^{(\underline{mn})}\,\delta(\mathbf{x}-\mathbf{x'})\, ,\\
\{A^{(\underline{m})a}_{\bar \mu}(x), \, \pi^{(\underline{n})b}_{\bar \nu}(x')\}&=\delta_{ab}\, \delta_{\bar \mu \bar  \nu}\, \delta^{(\underline{mn})}\,\delta(\mathbf{x}-\mathbf{x'})\, .
\end{align}
\end{subequations}
Here the symbol $\delta^{(\underline{mn})}$ stands for the following product of Kronecker's deltas:
\begin{equation}
\delta_{(\underline{r} \underline{s})}=\delta_{\underline{r}_{1}\, \underline{s}_{1}}\delta_{\underline{r}_{2}\, \underline{s}_{2}}\cdots\delta_{\underline{r}_{n}\, \underline{s}_{n}} \, .
\end{equation}
This shows that the transformations given by Eqs.~(\ref{MFV2}) and (\ref{MFS2}) correspond to a canonical map~\cite{OPT2}. It should be stressed that this result does not depend on the scheme used to carry out the compactification, it only depends on the completeness of the $\{f^{(\underline{m})}(\bar x)\}$ basis.
\\

The canonical map given by Eqs.~(\ref{MFV2}) and (\ref{MFS2}) clearly preserves the covariant essence of the first canonical map, Eq.~(\ref{MF1}), because it is immediate to establish that the $A^{(\underline{m})a}_\mu(x)$ and  $A^{(\underline{m})a}_{\bar \mu}(x)$ fields transform, under ${\rm SO}(1,3)$, as a vector and $n$ scalars, respectively. However, it is not clear how these objets transform under the standard ${\rm SU}(N,{\cal M}^4)$ gauge group; this point remains dark so far. To clarify it, note that the constant function $f^{(\underline{0})}$ may be an element of the set of functions $\{f^{(\underline{m})}(\bar x)\}$. From now on, we will assume that the constant function belongs to this set of complete functions. As already noted, according to map~(\ref{MF1}), the variation of the connections ${\cal A}^a_M(x,\bar x)$ given by Eq.~(\ref{GT}) unfold into
\begin{subequations}
\begin{align}
 \delta {\cal A}^a_\mu(x,\bar x)&={\cal D}^{ab}_\mu \alpha^b(x,\bar x) \, , \\
\delta {\cal A}^a_{\bar \mu}(x,\bar x)&={\cal D}^{ab}_{\bar \mu} \alpha^b(x,\bar x)\, .
\end{align}
\end{subequations}
Then, taking into account the canonical map given by Eqs.~(\ref{MFV2}) and (\ref{MFS2}), and using the orthogonality of the set $\{ f^{(\underline{0})},\, f^{(\underline{m})}(\bar x)\}$ of functions, we have
\begin{subequations}
\begin{align}
\label{GTV0}
\delta A^{(\underline{0})a}_\mu (x)&={\cal D}^{(\underline{0})ab}_\mu \alpha^{(\underline{0})b}(x)+gf^{abc}\sum_{(\underline{m})}A^{(\underline{m})b}_\mu(x)\alpha^{(\underline{m})c}(x) \, , \\
\label{GTVm}
\delta A^{(\underline{m})a}_\mu (x)&=gf^{abc} A^{(\underline{m})b}_\mu (x) \alpha^{(\underline{0})c}(x)+\sum_{(\underline{r})}{\cal D}^{(\underline{mr})ab}_\mu \alpha^{(\underline{r})b}(x) \, ,
\end{align}
\end{subequations}
\begin{subequations}
\begin{align}
\label{GTS0}
\delta A^{(\underline{0})a}_{\bar \mu} (x)&=gf^{abc}A^{(\underline{0})b}_{\bar \mu} \alpha^{(\underline{0})c}(x)+gf^{abc}\sum_{(\underline{m})}A^{(\underline{m})b}_{\bar \mu}(x)\alpha^{(\underline{m})c}(x)\nonumber \\
&+\sum_{(\underline{m})}\left[\int d^n\bar x f^{(\underline{0})}\partial_{\bar \mu} f^{(\underline{m})}\right]\alpha^{(\underline{m})a} \, , \\
\label{GTSm1}
\delta A^{(\underline{m})a}_{\bar \mu} (x)&=gf^{abc} A^{(\underline{m})b}_{\bar \mu} (x) \alpha^{(\underline{0})c}(x)+gf^{abc} A^{(\underline{0})b}_{\bar \mu} (x) \alpha^{(\underline{m})c}(x)\nonumber \\
&-\sum_{(\underline{r})}{\cal D}^{(\underline{mr})ab}_{\bar \mu} \alpha^{(\underline{r})b}(x) \, ,
\end{align}
\end{subequations}
where
\begin{subequations}
\begin{align}
\label{CDV}
{\cal D}^{(\underline{mr})ab}_\mu &=\delta^{(\underline{mr})}{\cal D}^{(\underline{0})ab}_\mu -gf^{abc}\sum_{(\underline{s})}\Delta_{(\underline{mrs})}A^{(\underline{s})c}_\mu  \, , \\
\label{CDS}
{\cal D}^{(\underline{mr})ab}_{\bar \mu} &=-\delta^{ab}\int d^n\bar x f^{(\underline{m})}(\bar x) \partial_{\bar \mu} f^{(\underline{r})}(\bar x) +gf^{abc}\sum_{(\underline{s})}\Delta_{(\underline{mrs})}A^{(\underline{s})c}_{\bar \mu } \, ,
\end{align}
\end{subequations}
with ${\cal D}^{(\underline{0})ab}_\mu=\delta^{ab}\partial_\mu-gf^{abc}A^{(\underline{0})c}_\mu$ the covariant derivative in the adjoint representation of ${\rm SU}(N,{\cal M}^4)$. In addition, to match the zero modes with those of the usual theory, the dimensionful coupling constant $g_{4+n}$ has been  rescaled to identify the 4-dimensional coupling $g=f^{(\underline{0})}g_{(4+n)}$. Further, the symbol $\Delta_{(\underline{mrs})}$ appearing in the above expressions is given by
\begin{equation}
\label{symbol}
\Delta_{(\underline{mrs})}=\frac{1}{f^{(\underline{0})}}\int d^n\bar x f^{(\underline{m})}(\bar x)f^{(\underline{r})}(\bar x)f^{(\underline{s})}(\bar x)\, ,
\end{equation}
which will be defined more precisely later. In previous expressions, generic quantities of the way $\varphi^{(\underline{0})}_A(x)$ and  $\varphi^{(\underline{m})}_A(x)$ are the components of fields or gauge parameters along $f^{(\underline{0})}$ and $f^{(\underline{m})}(x)$, respectively. Note that if we identify the $\alpha^{(\underline{0})a}(x)$ parameters as the gauge parameters of the standard ${\rm SU}(N,{\cal M}^4)$ group, then, making  $\alpha^{(\underline{m})a}(x)=0$, Eqs.~(\ref{GTV0}), (\ref{GTVm}), (\ref{GTS0}), and (\ref{GTSm1}) become
 \begin{subequations}
\begin{align}
\label{SGT0}
\delta_{\rm s} A^{(\underline{0})a}_\mu (x)&={\cal D}^{(\underline{0})ab}_\mu \alpha^{(\underline{0})b}(x) \, , \\
\label{SGTm}
\delta_{\rm s} A^{(\underline{m})a}_\mu (x)&=gf^{abc} A^{(\underline{m})b}_\mu (x) \alpha^{(\underline{0})c}(x)\, ,\\
\label{SGTS0}
\delta_{\rm s} A^{(\underline{0})a}_{\bar \mu} (x)&=gf^{abc}A^{(\underline{0})b}_{\bar \mu} \alpha^{(\underline{0})c}(x)\, , \\
\label{SGTSm}
\delta_{\rm s} A^{(\underline{m})a}_{\bar \mu} (x)&=gf^{abc} A^{(\underline{m})b}_{\bar \mu} (x) \alpha^{(\underline{0})c}(x) \, .
\end{align}
\end{subequations}
These expressions, which we call the SGTs, mean that $A^{(\underline{0})a}_\mu (x)$ transform as connections, whereas $A^{(\underline{m})a}_\mu (x)$, $A^{(\underline{0})a}_{\bar \mu} (x)$, and $A^{(\underline{m})a}_{\bar \mu} (x)$ transform as matter fields (in the adjoint representation) under this group. So far, everything seems to be going well. However, we must take into account the physical requirement that the standard pure Yang-Mills theory must be recovered in the limit as $R\to 0$. In other words, the new-physics effects must decouple, in accordance with the decoupling theorem~\cite{AC}. Nevertheless, the usual pure Yang-Mills theory does not have scalar fields, so the existence of such fields must be exclusively linked to the presence of extra dimensions. To establish this link, note that the only element of the set of basis functions which does not depend on the geometrical details of the compact manifold is the constant function  $f^{(\underline{0})}$. It is inferred from this that, in general, any field without standard counterpart does not have component along  $f^{(\underline{0})}$. The fact that the constant function is trivially even under the interchange $\bar x \to -\bar x$ suggests that the basis $\{f^{(\underline{0})}, \, f^{(\underline{m})}(\bar x)\}$ of functions can be reorganized into two independent bases, one containing the even functions, $\{f^{(\underline{0})}_{\rm E}, \, f^{(\underline{m})}_{\rm E}(\bar x)\}$, and the other containing the odd functions, $\{f^{(\underline{m})}_{\rm O}(\bar x)\}$. Then, we postulate that any field with standard counterpart is even under the reflection $\bar{x}\to-\bar{x}$, while those without standard counterpart are odd with respect to such a transformation. Accordingly, Eqs.(\ref{MFV2}, \ref{MFS2},\ref{MGP}) take the form
\begin{subequations}
\begin{align}
\label{MFV22}
{\cal A}^a_\mu(x,\bar x)&=f^{(\underline{0})}_{\rm E} A^{(\underline{0})a}_\mu(x)+\sum_{(\underline{m})}f^{(\underline{m})}_{\rm E}(\bar x)A^{(\underline{m})a}_\mu(x)\, , \\
\label{MFS22}
{\cal A}^a_{\bar \mu}(x,\bar x)&=\sum_{(\underline{m})}f^{(\underline{m})}_{\rm O}(\bar x)A^{(\underline{m})a}_{\bar \mu}(x)\, , \\
\label{MGP2}
\alpha^a(x,\bar x)&=f^{(\underline{0})}_{\rm E} \alpha^{(\underline{0})a}(x)+\sum_{(\underline{m})}f^{(\underline{m})}_{\rm E}(\bar x)\alpha^{(\underline{m})a}(x)\, ,
\end{align}
\end{subequations}
Also, this assumption eliminates Eqs.(\ref{GTS0},\ref{SGTS0}) and reduces Eq.(\ref{GTSm1}) to
\begin{equation}
\label{TGSm}
\delta A^{(\underline{m})a}_{\bar \mu} (x)=gf^{abc} A^{(\underline{m})b}_{\bar \mu}(x) \alpha^{(\underline{0})c}(x)-\sum_{(\underline{r})}{\cal D}^{(\underline{mr})ab}_{\bar \mu} \alpha^{(\underline{r})b}(x) \, ,
\end{equation}
while Eqs.~(\ref{CDV},\ref{CDS}) become
\begin{subequations}
\begin{align}
\label{CDV1}
{\cal D}^{(\underline{mr})ab}_\mu &=\delta^{(\underline{mr})}{\cal D}^{(\underline{0})ab}_\mu -gf^{abc}\sum_{(\underline{s})}\Delta_{(\underline{mrs})}A^{(\underline{s})c}_\mu  \, , \\
\label{CDS1}
{\cal D}^{(\underline{mr})ab}_{\bar \mu} &=-\delta^{ab}p^{(\underline{mr})}_{\bar \mu} +gf^{abc}\sum_{(\underline{s})}\Delta'_{(\underline{mrs})}A^{(\underline{s})c}_{\bar \mu } \, ,
\end{align}
\end{subequations}
where
\begin{equation}
\label{KE}
p^{(\underline{mr})}_{\bar \mu}=\int d^n\bar x f^{(\underline{m})}_{\rm O}(\bar x) \partial_{\bar \mu} f^{(\underline{r})}_{\rm E}(\bar x) \, .
\end{equation}
Observe that, due to the parity properties of the basis functions, the symbol given by Eq.(\ref{symbol}) splits into the following symbols:
\begin{subequations}
\begin{align}
\Delta_{(\underline{mrs})}&=\frac{1}{f^{(\underline{0})}}\int d^n\bar x f^{(\underline{m})}_{\rm E}(\bar x)f^{(\underline{r})}_{\rm E}(\bar x)f^{(\underline{s})}_{\rm E}(\bar x)\, , \\
\Delta'_{(\underline{mrs})}&=\frac{1}{f^{(\underline{0})}}\int d^n\bar x f^{(\underline{m})}_{\rm O}(\bar x)f^{(\underline{r})}_{\rm O}(\bar x)f^{(\underline{s})}_{\rm E}(\bar x)\, .
\end{align}
\end{subequations}
Directly from their definitions, we can see that $\Delta_{(\underline{mrs})}$ is symmetric in all its indices, while $\Delta'_{(\underline{mrs})}$ is symmetric only under the interchange of the first two indices. Note that the determination of these symbols requires the specifcation of the set of functions $f^{(\underline{m})}(\bar x)$.
\\

In order to identify the variations of the fields $A^{(\underline{0})a}_{ \mu}(x)$, $A^{(\underline{m})a}_{ \mu}(x)$, and $A^{(\underline{m})a}_{\bar \mu}(x)$ under the ${\rm SU}(N,{\cal M}^4)$ group, we have taken the parameters $\alpha^{(\underline{m})a}(x)$ equal to zero, but such parameters define transformations of these fields as well. This fact should not surprise us, as it is inherent to the implementation of a canonical transformation to hide a gauge symmetry. When the map connects two different Lie groups, that is, when it connects the gauge group $G$ to one of its subgroups $H$, the parameters analogous to $\alpha^{(\underline{m})a}(x)$ are those associated with the generators of $G$ which do not belong to $H$. In our case, the map is between two different gauge groups that are identical as Lie groups. However, in both cases, the presence of these parameters tells us that there is a larger gauge symmetry. The parameters $\alpha^{(\underline{m})a}(x)$ play a central role in quantizing the theory in the context of the field-antifield formalism. This problem has already been addressed by some of us in previous works~\cite{PRDYM,OPT1} based on the BRST symmetry~\cite{BRST}.\\

It should be stressed that the gauge structure of the original theory is not spoiled, since the algebra satisfied by the constrains of the theory is given in terms of Poisson's brackets~\cite{Dirac}, which are preserved by canonical maps (see Ref.~\cite{OPT2}). A direct consequence of this is the fact that the KK excitations $A^{(\underline{m})a}_\mu(x)$ are gauge fields, although from the ${\rm SU}(N,{\cal M}^4)$ perspective they transform as matter fields. The $\alpha^{(\underline{m})a}(x)$ parameters define the NSGTs according to
\begin{subequations}
\begin{align}
\label{NSGTV0}
\delta_{\rm ns} A^{(\underline{0})a}_\mu (x)&=gf^{abc}\sum_{(\underline{m})}A^{(\underline{m})b}_\mu(x)\alpha^{(\underline{m})c}(x) \, , \\
\label{NSGTVm}
\delta_{\rm ns} A^{(\underline{m})a}_\mu (x)&=\sum_{(\underline{r})}{\cal D}^{(\underline{mr})ab}_\mu \alpha^{(\underline{r})b}(x) \, ,\\
\label{GTSm}
\delta_{\rm ns} A^{(\underline{m})a}_{\bar \mu} (x)&=-\sum_{(\underline{r})}{\cal D}^{(\underline{mr})ab}_{\bar \mu} \alpha^{(\underline{r})b}(x) \, .
\end{align}
\end{subequations}
\\

So far, we have completed the program of hiding the ${\rm ISO}(1,3+n)\otimes{\rm SU}(N,{\cal M}^d)$ symmetry into the ${\rm ISO}(1,3)\otimes{\rm SU}(N,{\cal M}^4)$ symmetry through the canonical transformations (\ref{MF1}), (\ref{MFV22}), and (\ref{MFS22}), together with the map of the gauge parameters given in Eq.~(\ref{MGP2}). The original connections ${\cal A}^a_M(x,\bar x)$ have been mapped into the infinite set of fields $\{ A^{(\underline{0})a}_\mu(x), A^{(\underline{m})a}_\mu(x), A^{(\underline{m})a}_{\bar \mu}(x)\}$. We have shown that the $A^{(\underline{0})a}_\mu$ fields correspond to the connections of the gauge group ${\rm SU}(N,{\cal M}^4)$, whereas the infinite set of fields $\{A^{(\underline{m})a}_\mu, A^{(\underline{m})a}_{\bar \mu}\}$, which are recognized as the KK excitations of $A^{(\underline{0})a}_\mu$, transform in the adjoint representation of this group. As commented at the beginning of this section, when the gauge symmetry $G$ is hidden into $H$, with $H$ a subgroup of $G$, the connections of $G$, $A^a_\mu$, are mapped into the connections $A^{\bar a}_\mu$, and matter fields, $A^{\hat a}_\mu$, of $H$. From the $H$ perspective, the fields $A^{\hat a}_\mu$ can be endowed with mass, for which SSB can be used. In our case, the $\{A^{(\underline{m})a}_\mu, A^{(\underline{m})a}_{\bar \mu}\}$ fields appear as matter fields under the ${\rm SU}(N,{\cal M}^4)$ perspective, so they can be endowed with mass. However, not all these fields can be identified as gauge fields, since there must be a one-to-one relation between gauge fields and gauge parameters. It is not difficult to convince ourselves that the only possibility is $A^{(\underline{m})a}_\mu(x) \leftrightarrow \alpha^{(\underline{m})a}(x)$, so the $A^{(\underline{m})a}_{\bar \mu}$ fields are not gauge fields, as it was expected. At this stage, we can ask the following question: What is the instrument analogous to SSB that allows us to endow these fields with mass? We now answer this question.
\\

\subsection{Mass spectrum}
Previously, we have shown how to hide the ${\rm ISO}(1,3+n) \otimes {\rm SU}(N,{\cal M}^d)$ symmetry into the standard ${\rm ISO}(1,3) \otimes {\rm SU}(N,{\cal M}^4)$ symmetry through the canonical transformations (\ref{MF1}) and (\ref{MFV22},\ref{MFS22},\ref{MGP2}). The next step is to identify the mechanism that allows us to endow the gauge, $A^{(\underline{m})a}_\mu(x)$, and scalar, $A^{(\underline{m})a}_{\bar \mu}(x)$, fields with mass, proportional to the compactification scale $R^{-1}$. In this respect, we observe that such a mechanism is contained in the  very structure of the theory. To see this, note that the curvatures ${\cal F}^a_{\mu \nu}(x,\bar x)$, ${\cal F}^a_{\mu \bar \nu}(x,\bar x)$, and ${\cal F}^a_{\bar \mu \bar \nu}(x,\bar x)$ are $\bar{x}$ even, $\bar{x}$ odd, and $\bar{x}$ even, respectively, so they can be expressed in the $\{ f^{(\underline{0})}_{\rm E},\, f^{(\underline{m})}_{\rm E}(\bar x)\}$ and $\{f^{(\underline{m})}_{\rm O}(\bar x)\}$ bases as follows:
\begin{subequations}
\begin{align}
\label{MCVV}
{\cal F}^a_{\mu \nu}(x,\bar x)&=f^{(\underline{0})}_{\rm E}\,{\cal F}^{(\underline{0})a}_{\mu \nu}(x) +\sum_{(\underline{m})}f^{(\underline{m})}_{\rm E}(\bar x)\,{\cal F}^{(\underline{m})a}_{\mu \nu}(x)\, , \\
\label{MCVS}
{\cal F}^a_{\mu \bar \nu}(x,\bar x)&=\sum_{(\underline{m})}f^{(\underline{m})}_{\rm O}(\bar x)\,{\cal F}^{(\underline{m})a}_{\mu \bar \nu}(x)\, , \\
\label{MCSS}
{\cal F}^a_{\bar \mu \bar \nu}(x,\bar x)&=f^{(\underline{0})}_{\rm E}\,{\cal F}^{(\underline{0})a}_{\bar \mu \bar \nu}(x) +\sum_{(\underline{m})}f^{(\underline{m})}_{\rm E}(\bar x)\,{\cal F}^{(\underline{m})a}_{\bar \mu \bar \nu}(x)\, .
\end{align}
\end{subequations}
Then, after integrating over the extra coordinates in the action displayed in Eq.~(\ref{A2}), we have an effective Lagrangian given by
\begin{equation}
\label{EL}
 {\cal L}^{\rm YM}_{\rm eff}= {\cal L}^{\rm YM}_{\textrm{v-v}}+ {\cal L}^{\rm YM}_{\textrm{v-s}}+ {\cal L}^{\rm YM}_{\textrm{s-s}}\, ,
\end{equation}
with
\begin{subequations}
\begin{align}
\label{ELVV}
{\cal L}^{\rm YM}_{\textrm{v-v}}&=-\frac{1}{4}{\cal F}^{(\underline{0})a}_{\mu \nu}(x){\cal F}^{(\underline{0})\mu \nu}_a(x) -\frac{1}{4}
\sum_{(\underline{m})}{\cal F}^{(\underline{m})a}_{\mu \nu}(x){\cal F}^{(\underline{m})\mu \nu}_a(x)\, , \\
\label{ELVS}
{\cal L}^{\rm YM}_{\textrm{v-s}}&=\frac{1}{2}
\sum_{(\underline{m})}{\cal F}^{(\underline{m})a}_{\mu \bar \nu}(x){\cal F}^{(\underline{m})a\, \mu}\hspace{0.00001cm}_{\bar \nu}(x)\, , \\
\label{ELSS}
{\cal L}^{\rm YM}_{\textrm{s-s}}&=-\frac{1}{4}{\cal F}^{(\underline{0})a}_{\bar \mu \bar \nu}(x){\cal F}^{(\underline{0})\bar \mu \bar \nu}_a(x) -\frac{1}{4}
\sum_{(\underline{m})}{\cal F}^{(\underline{m})a}_{\bar \mu \bar \nu}(x){\cal F}^{(\underline{m})\bar \mu \bar \nu}_a(x)\, .
\end{align}
\end{subequations}
Note that the Lagrangians (\ref{ELVS}) and (\ref{ELSS}) correspond to a scalar kinetic sector and to a scalar potential, respectively. This means that  masses for the gauge, $A^{(\underline{m})a}_\mu(x)$, and scalar, $A^{(\underline{m})a}_{\bar \mu}(x)$, fields can arise from the Lagrangians (\ref{ELVS}) and (\ref{ELSS}), respectively. The presence of quadratic terms in (\ref{ELVS}) and (\ref{ELSS}) can be seen noting that the 1-form and 0-form that define  these Lagrangians have parts that are linear in fields:
\begin{subequations}
\begin{align}
\label{CV}
{\cal F}^{(\underline{m})a}_{\mu \bar \nu}&={\cal D}^{(\underline{0})ab}_\mu A^{(\underline{m})b}_{\bar \nu}-\sum_{(\underline{r})}A^{(\underline{r})a}_\mu\int d^n\bar xf^{(\underline{m})}_{\rm O}(\bar x)\partial_{\bar \nu}f^{(\underline{r})}_{\rm E}(\bar x)\nonumber \\
&+
gf^{abc}\sum_{(\underline{rs})}\Delta'_{(\underline{rsm})}A^{(\underline{r})b}_\mu (x)A^{(\underline{s})c}_{\bar \nu}(x)\\
\label{CS}
{\cal F}^{(\underline{m})a}_{\bar \mu \bar \nu}&=\sum_{(\underline{r})}A^{(\underline{r})a}_{\bar \alpha}\int d^n\bar x f^{(\underline{m})}_{\rm E}(\bar x)\Big[\delta_{\bar \alpha \bar \nu}\, \partial_{\bar \mu}f^{(\underline{r})}_{\rm O}(\bar x)-\delta_{\bar \alpha \bar \mu}\, \partial_{\bar \nu}f^{(\underline{r})}_{\rm O}(\bar x)\Big]\nonumber \\
&+gf^{abc}\sum_{(\underline{rs})}\Delta'_{(\underline{rsm})}A^{(\underline{r})b}_{\bar \mu}(x)A^{(\underline{s})c}_{\bar \nu}(x) \, ,
\end{align}
\end{subequations}
\begin{equation}
\label{CSF0}
{\cal F}^{(\underline{0})a}_{\bar \mu \bar \nu}=gf^{abc}\sum_{(\underline{m})}A^{(\underline{m})b}_{\bar \mu}(x)A^{(\underline{m})c}_{\bar \nu}(x)\, .
\end{equation}
The structures of Eqs.~(\ref{CV}) and (\ref{CS}) are quite illuminating. Firstly, note that the functions $\partial_{\bar \mu}f^{(\underline{m})}_{\rm E}(\bar x)$ and $\partial_{\bar \mu}f^{(\underline{m})}_{\rm O}(\bar x)$ are odd and even, respectively. This fact guarantees the existence of mass terms for the fields $A^{(\underline{m})a}_\mu(x)$ and $A^{(\underline{m})a}_{\bar \mu}(x)$, and suggests how to define the set of orthonormal functions $\{f^{(\underline{0})}_{\rm E}, \, f^{(\underline{m})}_{\rm E}(\bar x), \, f^{(\underline{m})}_{\rm O}(\bar x) \}$.
\\

\section{The set of orthonormal functions}
\label{OF}
In the previous section, we have emphasized the crucial role played by the complete set of orthonormal functions $\{f^{(\underline{0})}_{\rm E}, \, f^{(\underline{m})}_{\rm E}(\bar x), \, f^{(\underline{m})}_{\rm O}(\bar x) \}$ to go from the description based on ${\rm ISO}(1,3+n) \otimes {\rm SU}(N,{\cal M}^d)$ to the description governed by ${\rm ISO}(1,3) \otimes {\rm SU}(N,{\cal M}^4)$, using even and odd series. We can highlight the importance of this canonical map by noting that: $(1)$ it allows us to identify the degrees of freedom of the four-dimensional theory as the coefficients of these infinite series; $(2)$ it allows us to establish whether a given KK excitation is a gauge field or a matter field of ${\rm SU}(N,{\cal M}^4)$; and $(3)$ it allows us to render the KK excitations massive.
\\

To generate the complete set of functions $\{f^{(\underline{0})}_{\rm E}, \, f^{(\underline{m})}_{\rm E}(\bar x), \, f^{(\underline{m})}_{\rm O}(\bar x) \}$, we need some observable that can be associated with extra dimensions. By observable we mean a Hermitian operator that allows us to generate an orthogonal basis which is  associated to a set of real eigenvalues. The eigenvalues of this observable must correspond to the masses of the KK fields. In this respect, we note that we have at our disposal the Casimir invariants~\cite{Cha} of the inhomogeneous ${\rm ISO}(n)$ group, whose generators, as it was shown in Sec.~\ref{P}, are all constants of motion. One of these invariants is the one associated with the translations group ${\rm T}(n)$, given by ${\bar P}^2= P_{\bar \mu} { P}_{\bar \mu}$. The observables $P_{\bar \mu}$ and ${\bar P}^2$ generate a basis of simultaneous eigenkets $\{ |\bar p \big>\}$ through the fundamental equations:
\begin{subequations}
\begin{align}
\label{ev1}
P_{\bar \mu}|\bar p\big >&=p_{\bar \mu}|\bar p\big > \, ,\\
\label{ev2}
\bar{P}^2 |\bar p\big >&=\bar{p}^2|\bar p\big >\, ,
\end{align}
\end{subequations}
where $\bar{p}^2\equiv p_{\bar \mu} p_{\bar \mu}$. The zero eigenvalue $p_{\bar \mu}=0$ is the link that allows us to make contact with the usual Yang-Mills theory. In general, the spectrum of eigenvalues can be continuous, but this can change when we make contact with the compact manifold. We assume that the basis $\{ |\bar p \big>\}$ is normalized and complete, that is, $\big<\bar p'|\bar p\big >=\delta_{\bar p' \bar p}$ and $\sum_{\bar p}|\bar p\big >\big<\bar p|=\mathbb{I}$. To make contact with the compact manifold, we need to represent Eqs.(\ref{ev1},\ref{ev2}) in the coordinates basis. Let $\bar X^\dag=\bar X$ be the Hermitian position operator, which satisfies the commutation relation $[X_{\bar \mu},P_{\bar \nu}]=i \delta_{\bar \mu \bar \nu}$. This operator generates the position basis $\{ |\bar x \big> \}$ through the eigenvalues equation
\begin{equation}
X_{\bar \mu}|\bar x \big>=x_{\bar \mu}|\bar x \big>\, ,
\end{equation}
which is orthonormal and complete in the Dirac's sense:
\begin{subequations}
\begin{align}
&\big < \bar x'|\bar x\big>=\delta(\bar x'-\bar x)\, ,\\
&\int d^n\bar x |\bar x \big>\big<\bar x|=\mathbb{I}\, .
\end{align}
\end{subequations}
En this basis $P_{\bar \mu}\to -i\partial_{\bar \mu}$, so Eq.~(\ref{ev2}) becomes
\begin{equation}
\left(\bar{\nabla}^2+\bar{p}^2 \right)f_{\bar p}(\bar x)=0\, ,
\end{equation}
where $f_{\bar p}(\bar x)=\big< \bar x| \bar p \big>$. Clearly, the solutions of this equation are plane waves $\exp(i\bar p\cdot \bar x)$, that is, a linear combination of $n$-dimensional sines and cosines. In an infinite space, the spectrum is continuous, but our manifold is compact. It is time to define the geometry of our compact manifold. We assume that each coordinate $\bar x_i$ is coiled in a circle $S^1$ of radius $R_i$, which are assumed, for simplicity, to be all equal, namely $R_1=\cdots=R_n\equiv R$. So, up to this stage, our compact manifold is the direct product of $n$ circles. This geometry introduces periodicity in our fields, that is, all our fields satisfy $\varphi_a(x,\bar x+ 2\pi R)=\varphi_a(x,\bar x )$. This in turn implies that the series introduced previously are actually general Fourier series, that is, our set of base functions are multidimensional sines and cosines. However, we have already seen that to make contact with the usual Yang-Mills theory, we need to introduce a definite parity for all fields $\varphi_a(x,\bar x )$ under the operation $\bar x \to -\bar x$. To implement this symmetry we introduce the cyclic group $Z_2$, which assigns to every point $\bar x$ of each circle its antipode $-\bar x$. Introducing this symmetry turns each circle $S^1$ into the orbifold $S^1/Z_2$. Then, we assume a compact manifold made of $n$ copies of the $S^1/Z_2$ orbifold. Then, we demand that the odd functions (sines) satisfy Dirichlet's boundary conditions and that the even functions (cosines) satisfy Neumann's boundary conditions, that is,
\begin{equation}
\sin\left(p_{\bar \mu} x_{\bar \mu}\right)\bigg |^{2\pi R}_0=0\, , \, \, \, \partial_{\bar \mu}\left[\cos\left(p_{\bar \mu} x_{\bar \mu}\right)\right]\bigg |^{2\pi R}_0=0\, .
\end{equation}
These boundary conditions lead to a discrete spectrum for the set of eigenvalues $p_{\bar \mu}$, that is, $p^{(\underline{m})}_{\bar \mu}=(\frac{\underline{m}}{R})$. This is a vector with $n$ entries, with at least one non-zero, but where all possibilities are considered (see Appendix \ref{S} for notation). The null eigenvalue $p_{\bar \mu}=0$ will be considered separately, since it plays an important role in defining the standard fields (the zero mode fields). Then, the basis generated by the observables $P_{\bar \mu}$ and $\bar{P}^2$ is discrete, which we thus denote by $\{ |0\big>, |\bar{p}^{(\underline{m})}\big>\}$. The corresponding eigenfunctions are given by $f^{(\underline{0})}_{\rm E}=\big<\bar x|0\big >$ and $f^{(\underline{m})}_{\rm E}(\bar x), f^{(\underline{m})}_{\rm O}(\bar x)=\big<\bar x|\bar{p}^{(\underline{m})}\big>$. So, our complete set of orthonormal functions $\{ f^{(\underline{0})}_{\rm E}, f^{(\underline{m})}_{\rm E}(\bar x), f^{(\underline{m})}_{\rm O}(\bar x)\}$, introduced in the subsection \ref{hs}, corresponds to the set of eigenfunctions associated with the mass eigenvalues of $\bar{P}^2$, given by $p^{(\underline{m})}_{\bar \mu}p^{(\underline{m})}_{\bar \mu}\equiv m^2_{(\underline{m})}=\frac{\underline{m}^2}{R^2}$, with $f^{(\underline{0})}_{\rm E}$ corresponding to the eigenvalue  $p^{(\underline{m})}_{\bar \mu}p^{(\underline{m})}_{\bar \mu}=0$. The fact that the eigenfunction $f^{(\underline{0})}_{\rm E}$ is associated with an eigenvalue equal to zero means that the standard fields do not receive mass at the $R^{-1}$ scale. Accordingly, the set of maps (\ref{MFV22}, \ref{MFS22}, \ref{MGP2}) and (\ref{MCVV}, \ref{MCVS}, \ref{MCSS}) are given by even or odd $n$-dimensional Fourier series. The set of functions $\{f^{(\underline{0})}_{\rm E},\,f^{(\underline{m})}_{\rm E}(\bar x), \, f^{(\underline{m})}_{\rm O}(\bar x)\}$ is given in Appendix ~\ref{AE}.
\\

With the help of the set $\{f^{(\underline{0})}_{\rm E},\,f^{(\underline{m})}_{\rm E}(\bar x), \, f^{(\underline{m})}_{\rm O}(\bar x)\}$ shown in Appendix~\ref{AE}, we can rewrite the expressions (\ref{CV}) and (\ref{CS}) as follows:
\begin{subequations}
\begin{align}
\label{CVF}
{\cal F}^{(\underline{m})a}_{\mu \bar \nu}&={\cal D}^{(\underline{0})ab}_\mu A^{(\underline{m})b}_{\bar \nu}(x)+p^{(\underline{m})}_{\bar \nu}A^{(\underline{m})a}_{ \mu}(x)\nonumber \\
&+gf^{abc}\sum_{(\underline{rs})}\Delta'_{(\underline{rsm})}A^{(\underline{r})b}_\mu(x)A^{(\underline{s})c}_{\bar \nu}(x)\, , \\
\label{CSF}
{\cal F}^{(\underline{m})a}_{\bar \mu \bar \nu}&=p^{(\underline{m})}_{\bar \mu}A^{(\underline{m})a}_{\bar \nu}-p^{(\underline{m})}_{\bar \nu}A^{(\underline{m})a}_{\bar \mu}+gf^{abc}\sum_{(\underline{rs})}\Delta'_{(\underline{rsm})}A^{(\underline{r})b}_{\bar \mu}A^{(\underline{s})c}_{\bar \nu}\, ,
\end{align}
\end{subequations}
where
\begin{equation}
p_{\bar{\mu}}^{(\underline m)} =\sum^n_{\alpha=1}\frac{\underline{m}_\alpha}{R_\alpha}\delta_{\bar{\mu}\, {4+\alpha}}\, ,
\end{equation}
\begin{equation}
\label{MassA}
m^2_{(\underline{m})}=\left(\frac{ \underline{m}_1}{R_1}\right)^2+\cdots +\left(\frac{\underline{m}_n}{R_n}\right)^2=p_{\bar{\mu}}^{(\underline m)}  p_{\bar{\mu}}^{(\underline m)} \, .
\end{equation}
Recall that we are assuming that all radii $R_i$ are equal. On the other hand, the 2-form curvature components $\{{\cal F}^{(\underline{0})a}_{\mu \nu}(x), {\cal F}^{(\underline{m})a}_{\mu \nu}(x)\}$ are given by
\begin{subequations}
\begin{align}
\label{CVV0}
{\cal F}^{(\underline{0})a}_{\mu \nu}(x)&=F^{(\underline{0})a}_{\mu \nu}(x)+gf^{abc}\sum_{(\underline{m})}A^{(\underline{m})b}_\mu(x)A^{(\underline{m})c}_\nu(x)\, ,\\
\label{CVVm}
{\cal F}^{(\underline{m})a}_{\mu \nu}(x)&={\cal D}^{(\underline{0})ab}_\mu A^{(\underline{m})b}_\nu(x)-{\cal D}^{(\underline{0})ab}_\nu A^{(\underline{m})b}_\mu(x)\nonumber \\
&+gf^{abc}\sum_{(\underline{rs})}\Delta_{(\underline{mrs})}A^{(\underline{r})b}_\mu(x)A^{(\underline{s})c}_\nu(x)\, ,
\end{align}
\end{subequations}
where
\begin{equation}
\label{CVV0}
F^{(\underline{0})a}_{\mu \nu}(x)=\partial_\mu A^{(\underline{0})a}_\nu(x)-\partial_\nu A^{(\underline{0})a}_\mu(x)+gf^{abc}A^{(\underline{0})b}_\mu(x)A^{(\underline{0})c}_\nu(x)\, ,
\end{equation}
are the curvature components associated with the standard gauge group ${\rm SU}(N,{\cal M}^4)$. Note that the right-hand side of Eqs.~(\ref{CVVm}) and (\ref{CVV0}) transform in the adjoint representation of this group, as it must be. Then, the Lagrangian (\ref{ELVV}) contains the Yang-Mills term associated with the standard ${\rm SU}(N,{\cal M}^4)$ group plus terms involving interactions among connection components and matter fields. Note that ${\cal F}^{(\underline{m})a}_{\mu \nu}(x)$ contains the kinetic terms for the matter fields. This Lagrangian resembles the one emerging from the Yang-Mills sector of the electroweak group. In this case, the role played by the  ${\rm SU}(N,{\cal M}^4)$  group is analogous to the role played by the electromagnetic group ${\rm U}_e(1,{\cal M}^4)$ emerged from the implementation of SSB to the electroweak sector of the SM, whereas the role played by the matter fields $A^{(\underline{m})a}_\mu$ evokes that of the $W^{(\underline{0})\pm}_\mu$ bosons.
\\

As far as the mass terms for the KK $A^{(\underline{m})a}_\mu(x)$ fields are concerned, using the expression for the 1-form curvature, Eq.~ (\ref{CVF}), the Lagrangian (\ref{ELVS}) becomes
\begin{eqnarray}
\label{ELVS1}
{\cal L}^{\rm YM}_{\textrm{v-s}}&=&\sum_{(\underline{m})}\Big[\frac{1}{2}\left({\cal D}^{(\underline{0})ab}_\mu A^{(\underline{m})b}_{\bar \nu}\right)\left({\cal D}^{(\underline{0})ac\mu} A^{(\underline{m})c}_{\bar \nu}\right)\nonumber \\
&&+p^{(\underline{m})}_{\bar \nu}A^{(\underline{m})a\mu}\left({\cal D}^{(\underline{0})ab}_\mu A^{(\underline{m})b}_{\bar \nu}\right)+\frac{1}{2}m^2_{(\underline{m})} A^{(\underline{m})a}_\mu A^{(\underline{m})a\mu}   \Big]\nonumber  \\
&&+gf^{abc}\sum_{(\underline{mrs})}\Delta'_{(\underline{rsm})}\Big[\left({\cal D}^{(\underline{0})ad}_\mu A^{(\underline{m})d}_{\bar \nu}\right)A^{(\underline{r})b\mu}A^{(\underline{s})c}_{\bar \nu}\nonumber \\
&&+\frac{g}{2}f^{ade}\sum_{(\underline{pq})}\Delta'_{(\underline{pqm})}A^{(\underline{r})b}_\mu A^{(\underline{s})c}_{\bar \nu} A^{(\underline{p})d\mu} A^{(\underline{q})e}_{\bar \nu}\Big]\, .
\end{eqnarray}
Observe the resemblance of this term with a Higgs kinetic term. In fact, it has the same gauge and Lorentz structures of these kind of terms. In particular, we would like to highlight the presence of mass terms for the KK vector $A^{(\underline{m})a}_\mu$ fields and also the presence of bilinear and trilinear interactions proportional to the $p^{(\underline{m})}_{\bar \nu}$ scale, which are essential ingredients of a Higgs kinetic term.
\\

From the term proportional to $ \mathcal{F}_{\bar{\mu}\bar{\nu}}^{(\underline{m})a} \mathcal{F}^{(\underline{m})\bar{\mu}\bar{\nu}}_{a}$, in the Lagrangian \eqref{ELSS}, massive scalars and massless scalars (pseudo--Goldstone bosons) arise. In the general case, the scalar fields $ A^{(\underline{m})a}_{\bar{\mu}} $ show themselves in bilinear forms, so that, after proper diagonalizations, one recognizes a pseudo-Goldstone boson, which we denote by $ A^{(\underline{m})a}_{\rm G} $, and $n-1$ massive scalar fields, represented here by $ A^{(\underline{m})a}_{\bar{n}} $, with $ \bar{n}=1,2,\ldots,n-1 $. Using the above expressions for the 0-form curvature components, it can be seen that
\begin{eqnarray}
\label{ELSS1}
{\cal L}^{\rm YM}_{\textrm{s-s}}&=&-\sum_{(\underline{m})} \Bigg \{ \frac{1}{2}A^{(\underline{m})a}_{\bar \mu}\mathfrak{M}_{\bar \mu \bar \nu }^{(\underline{m})}A^{(\underline{m})a}_{\bar \nu}+gp^{(\underline{m})}_{\bar \mu}\sum_{(\underline{rs})}\Delta'_{(\underline{rsm})}\left(f^{abc}A^{(\underline{m})a}_{\bar \nu}A^{(\underline{r})b}_{\bar \mu}A^{(\underline{s})c}_{\bar \nu}\right)\nonumber \\
&&+\frac{g^2}{4}\sum_{(\underline{r})}\Big[\left(f^{eab}A^{(\underline{m})a}_{\bar \mu} A^{(\underline{m})b}_{\bar \nu}\right) \left(f^{ecd}A^{(\underline{r})c}_{\bar \mu} A^{(\underline{r})d}_{\bar \nu}\right)\nonumber \\
&&+\sum_{(\underline{spq})}\Delta'_{(\underline{rsm})} \Delta'_{(\underline{pqm})}\left(f^{eab}A^{(\underline{r})a}_{\bar \mu} A^{(\underline{s})b}_{\bar \nu}\right)\left(f^{ecd}A^{(\underline{p})c}_{\bar \mu} A^{(\underline{q})d}_{\bar \nu}\right)\Big]
\Bigg \},
\end{eqnarray}
where, for each possible value of  $ (\underline{m}) $, the corresponding $ n\times n $ symmetric mass matrix is
\begin{equation}
\label{MM}
\mathfrak{M}^{(\underline{m})}_{\bar{\mu}\bar{\nu}}=m_{(\underline{m})}^{2}\delta_{\bar{\mu}\bar{\nu}}-p_{\bar{\mu}}^{(\underline m)}  p_{\bar{\nu}}^{(\underline m)}\, .
\end{equation}
This matrix, whose mathematical structure is directly dictated by gauge invariance, naturally leads to the existence of a massless scalar field. To see this, we take the trace in Eq. (\ref{MM}) to obtain
\begin{equation}
\delta^{\bar \mu  \bar \nu}\mathfrak{M}^{(\underline{m})}_{\bar{\mu}\bar{\nu}}=(n-1)m^2_{(\underline{m})}\, ,
\end{equation}
which shows that there exist $(n-1)$ mass-degenerate physical scalar fields and a massless scalar field, which can be identified with a pseudo-Goldstone boson. This matrix can easily be diagonalized by noting that it coincides with the inertia tensor, per mass unit, of a single massive particle located at $\textrm{\textbf{r}}^{\dagger}=(p_{5},\ldots,p_{n+4}) $, with respect to some $n$-dimensional Euclidean reference system, rotating with angular velocity $ \omega $ about an arbitrary axis $ \hat{\boldsymbol{\omega}}=\boldsymbol{\omega}/\omega $. Hence, $ \mathfrak{M}^{(\underline{m})}_{\bar{\mu}\bar{\nu}} $ can be decomposed as
\begin{equation}\label{MD}
\mathfrak{M}^{(\underline{m})}_{\bar{\mu}\bar{\nu}}=
\mathcal{R}^{(\underline{m})}_{\bar{\mu}\bar{\mu}'}\mathfrak{M}^{(\underline{m})}_{\bar{\mu}'\bar{\nu}'}
\mathcal{R}^{(\underline{m})}_{\bar{\nu}\bar{\nu}'}\ .
\end{equation}
Here $ \left(\mathfrak{M}^{(\underline{m})}_{\bar{\mu}'\bar{\nu}'}\right)=\textrm{diag}(m_{(\underline{m})}^{2},m_{(\underline{m})}^{2},\ldots,m_{(\underline{m})}^{2},0) $, and $ \mathcal{R}^{(\underline{m})}=\left(\mathcal{R}^{(\underline{m})}_{\bar{\mu}\bar{\mu}'}\right) $ is an orthogonal matrix that transforms the mass-eigenvector components $ A_{\bar{\mu}'}^{(\underline{m})a}\equiv ( A_{\bar{n}}^{(\underline{m})a},  A_{\rm G}^{(\underline{m})a}) $ into $  A_{\bar{\mu}}^{(\underline{m})a} $, that is,
\begin{equation}\label{RT}
 A_{\bar{\mu}}^{(\underline{m})a}=\mathcal{R}^{(\underline{m})}_{\bar{\mu}\bar{\mu}'} A_{\bar{\mu}'}^{(\underline{m})a}=\mathcal{R}^{(\underline{m})}_{\bar{\mu}\bar{n}} A_{\bar{n}}^{(\underline{m})a}+\mathcal{R}^{(\underline{m})}_{\bar{\mu}{\rm G}} A_{\rm G}^{(\underline{m})a},\quad\bar{n}=1,2,\ldots,n-1\ ,
\end{equation}
where
\begin{equation}\label{RR}
\mathcal{R}^{(\underline{m})}_{\bar{\mu}\bar{\mu}'}\mathcal{R}^{(\underline{m})}_{\bar{\nu}\bar{\mu}'}=
\delta_{\bar{\mu}\bar{\nu}} \ \ \textrm{and}\ \ \ \mathcal{R}^{(\underline{m})}_{\bar{\mu}\bar{\mu}'}\mathcal{R}^{(\underline{m})}_{\bar{\mu}\bar{\nu}'}=
\delta_{\bar{\mu}'\bar{\nu}'}\ .
\end{equation}
In the above expressions, the primed indices denote mass eigenstates.\\ 

On the other hand, since $ \mathcal{R}^{(\underline{m})} $ is an orthogonal matrix with columns given by the principal-axes components, the useful identity
\begin{equation}\label{mR}
p_{\bar{\nu}}^{(\underline m)} \mathcal{R}^{(\underline{m})}_{\bar{\nu}\bar{\nu}'}=m_{(\underline{m})}\delta_{\bar{\nu}'{\rm G}}\
\end{equation}
follows. From this equation, together with Eq.(\ref{MassA}), one obtains
\begin{equation}
\label{MRI}
\mathcal{R}^{(\underline{m})}_{\bar{\nu}{\rm G}}=\frac{p_{\bar{\nu}}^{(\underline m)} } {m_{(\underline{m})}}\ .
\end{equation}
\\

Observe that neither $ \mathcal{F}^{(\underline{m})}_{\bar{\mu}\bar{\nu}} $ nor $ \mathcal{F}^{(\underline{0})}_{\bar{\mu}\bar{\nu}} $ follow a covariant transformation rule under $ \mathcal{R}^{(\underline{m})} $. Hence, once expressed in terms of the mass eigenvectors, Eq.~\eqref{ELSS1} explicitly depends on $ \mathcal{R}^{(\underline{m})} $. In fact, inserting Eqs.~\eqref{MD} and \eqref{RT} into Eq.~\eqref{ELSS1} we obtain
\begin{eqnarray}
{\cal L}^{\rm YM}_{\textrm{s-s}}&=&-\sum_{(\underline{m})}\Bigg\{ \frac{1}{2}m^2_{(\underline{m})}A^{(\underline{m})a}_{\bar n}A^{(\underline{m})a}_{\bar n}\nonumber \\
&&+gp^{(\underline{m})}_{\bar \mu}{\cal R}^{(\underline{m})}_{\bar \nu \bar \nu'}\sum_{(\underline{rs})}\Delta'_{(\underline{rsm})}{\cal R}^{(\underline{r})}_{\bar \mu \bar \mu'}{\cal R}^{(\underline{s})}_{\bar \nu \bar \lambda'}\left(f^{abc}A^{(\underline{m})a}_{\bar \nu'}A^{(\underline{r})b}_{\bar \mu'} A^{(\underline{s})c}_{\bar \lambda'}\right)\nonumber \\
&&+\frac{g^2}{4}\sum_{(\underline{r})}\Big[{\cal R}^{(\underline{m})}_{\bar \mu \bar \mu'}{\cal R}^{(\underline{m})}_{\bar \nu \bar \nu'}{\cal R}^{(\underline{r})}_{\bar \mu \bar \lambda'}{\cal R}^{(\underline{r})}_{\bar \nu \bar \rho'} \left(f^{eab}A^{(\underline{m})a}_{\bar \mu'} A^{(\underline{m})b}_{\bar \nu'}  \right)\left(f^{ecd}A^{(\underline{r})c}_{\bar \lambda'} A^{(\underline{r})d}_{\bar \rho'}  \right) \nonumber \\
&&+\sum_{(\underline{spq})}\Delta'_{(\underline{rsm})}\Delta'_{(\underline{pqm})}{\cal R}^{(\underline{r})}_{\bar \mu \bar \mu'}{\cal R}^{(\underline{s})}_{\bar \nu \bar \nu'}{\cal R}^{(\underline{p})}_{\bar \mu \bar \lambda'}{\cal R}^{(\underline{q})}_{\bar \nu \bar \rho'}\nonumber \\
&\times &\left(f^{eab}A^{(\underline{r})a}_{\bar \mu'} A^{(\underline{s})b}_{\bar \nu'}  \right)\left(f^{ecd}A^{(\underline{p})c}_{\bar \lambda'} A^{(\underline{q})d}_{\bar \rho'}  \right)
\Big]
\Bigg\}\, .
\end{eqnarray}
From this Lagrangian we immediately recognize the massive fields $ A^{(\underline{m})a}_{\bar{n}} $, with mass $ m_{(\underline{m})} $, and various trilinear- and quartic-interaction terms among them. It is worth emphasizing that all the mass terms are generated by the curvature components ${\cal F}^{(\underline{m})a}_{{\bar \mu}\nu}(x)$ and ${\cal F}^{(\underline{m})a}_{{\bar \mu}{\bar \nu}}(x)$, which in turn come from the extra-dimensional curvature ${\cal F}^a_{MN}(x,\bar x)$, whose precise structure is fixed by gauge invariance. So all masses originate in gauge symmetry and, in that sense, they can be properly called ``gauge masses''. As far as the massless scalar $ A^{(\underline{m})a}_{\rm G} $ fields are concerned, they play the role of pseudo-Goldstone bosons, as they can be removed from the theory through some sort of unitary gauge. The degrees of freedom that they represent appear as the longitudinal polarization states of the vector KK excitations $ A^{(\underline{m})a}_{\mu}$. The implementation of such unitary gauge can be understood in terms of the NSGTs. To see this, consider a specific NSGT with infinitesimal gauge parameters given by
\begin{equation}
\alpha^{(\underline{m})a}=\frac{ A^{(\underline{m})a}_{\rm G}}{m_{(\underline{m})}}\, .
\end{equation}
Then, at first order in the fields, the NSGTs given by Eqs. (\ref{NSGTVm}) and (\ref{GTSm}) become
\begin{subequations}
\begin{align}
A^{'(\underline{m})a}_{\mu}&=A^{(\underline{m})a}_{\mu}+\frac{\partial_\mu A^{(\underline{m})a}_{\rm G}}{m_{(\underline{m})}} \, , \\
A^{'(\underline{m})a}_{\bar{n}} &=A^{(\underline{m})a}_{\bar{n}} \, ,\\
A^{'(\underline{m})a}_{\rm G}&=0\, .
\end{align}
\end{subequations}
where use of relations (\ref{RT}) and (\ref{MRI}) was done. This is what we called the unitary gauge, due to its similarity with the gauge used in the SM in which all pseudo-Goldstone bosons are set to zero. In such a gauge one has $ A^{(\underline{m})a}_{\bar{\mu}} = \mathcal{R}^{(\underline{m})}_{\bar{\mu}\bar{\mu}'}A_{\bar{\mu}'}^{(\underline{m})a}= \mathcal{R}^{(\underline{m})}_{\bar{\mu}\bar{n}}A_{\bar{n}}^{(\underline{m})a}$. Note that the ${\cal L}^{\rm YM}_{\textrm{s-s}}$ Lagrangian plays the role of a Higgs potential, as it allows us to determine the presence of pseudo-Goldstone bosons and physical scalars.
\\

The Lagrangian (\ref{ELVS1}) is also affected by the orthogonal transformation (\ref{RT}). One has,
\begin{eqnarray}
\label{ELVS2}
{\cal L}^{\rm YM}_{\textrm{v-s}}&=&\sum_{(\underline{m})}\Bigg\{\frac{1}{2}\left({\cal D}^{(\underline{0})ab}_\mu A^{(\underline{m})b}_{\bar n}\right)\left({\cal D}^{(\underline{0})ac\mu} A^{(\underline{m})c}_{\bar n}\right)+\frac{1}{2}\left({\cal D}^{(\underline{0})ab}_\mu A^{(\underline{m})b}_{\rm G}\right)\left({\cal D}^{(\underline{0})ac\mu} A^{(\underline{m})c}_{\rm G}\right)\nonumber \\
&&+m_{(\underline{m})}A^{(\underline{m})a\mu}\left({\cal D}^{(\underline{0})ab}_\mu A^{(\underline{m})b}_{\rm G}\right)+\frac{1}{2}m^2_{(\underline{m})} A^{(\underline{m})a}_\mu A^{(\underline{m})a\mu}\nonumber \\
 &&+gf^{abc}\sum_{(\underline{rs})}\Delta'_{(\underline{rsm})}\Big[{\cal R}^{(\underline{m})}_{\bar \nu \bar \nu'}{\cal R}^{(\underline{s})}_{\bar \nu \bar \mu'}\left({\cal D}^{(\underline{0})ad}_\mu A^{(\underline{m})d}_{\bar \nu'}\right)A^{(\underline{r})b\mu}A^{(\underline{s})c}_{\bar \mu'}\nonumber \\
&&+\frac{g}{2}f^{ade}\sum_{(\underline{pq})}\Delta'_{(\underline{pqm})}{\cal R}^{(\underline{s})}_{\bar \nu \bar \mu'}{\cal R}^{(\underline{q})}_{\bar \nu \bar \nu'}A^{(\underline{r})b}_\mu A^{(\underline{s})c}_{\bar \mu'} A^{(\underline{p})d\mu} A^{(\underline{q})e}_{\bar \nu'}\Big]
\Bigg\}\, ,
\end{eqnarray}
where the relation (\ref{MRI}) was used. From the term proportional to the $m_{(\underline{m})}$ scale in the above expression, one can identify the presence of bilinear interactions between the vector KK excitations $A^{(\underline{m})a}_\mu$ and their associated pseudo-Goldstone bosons $A^{(\underline{m})a}_{\rm G}$. This class of interactions are typical of Higgs kinetic terms. \\

The canonical transformation given through the complete basis of functions $\{f^{(\underline{0})},\, f^{(\underline{m})}(\bar x)\}$ is enough to hide the  ${\rm ISO}(1,3+n) \otimes {\rm SU}(N,{\cal M}^d)$ symmetry into ${\rm ISO}(1,3) \otimes {\rm SU}(N,{\cal M}^4)$, since it allows us to establish the covariant character of the new fields under both the Poincar\' e group and the gauge group. The constant function $f^{(\underline{0})}_{\rm E}$ (the base state $|0\big>$ of $\bar{P}^2$) plays a central role in identifying the ${\rm SU}(N,{\cal M}^4)$ gauge parameters and connections. The choice of the $\bar{P}^2$ observable is very important because it allows us to endow the fields along the $f^{(\underline{m})}(\bar x)$ directions with mass. The split of the $\{f^{(\underline{0})},\, f^{(\underline{m})}(\bar x)\}$ basis into the sub-bases $\{f^{(\underline{0})}_E,\, f^{(\underline{m})}_{\rm E}(\bar x)\}$ and $\{f^{(\underline{m})}_{\rm O}(\bar x)\}$ is done in order to decouple effects of extra dimensions from the known physics. In particular, fields along $f^{(\underline{0})}$ cannot receive mass at the compactification scale $R^{-1}$, since the KK mechanism works through derivatives, with respect to the $\bar x$ coordinates, of the set $\{f^{(\underline{0})},\, f^{(\underline{m})}(\bar x)\}$ of functions. The choice of the Casimir invariant associated with the subgroup of the translations of the inhomogeneous group ${\rm ISO}(n)$ is natural,  given the structure of the curvatures ${\cal F}^{(\underline{m})a}_{\mu \bar \nu}(x,\bar x)$ and ${\cal F}^{(\underline{m})a}_{\bar \mu \bar \nu}(x,\bar x)$. After all, this invariant has to do with the homogeneity of the flat ${\cal N}^n$ space. Note that for $n=1$ there are no physical scalars, since in this case the curvatures ${\cal F}^{(\underline{m})a}_{\bar \mu \bar \nu}(x,\bar x)$ are absent. It is important to emphasize the fact that in order to match the zero modes with those of the usual theory, the dimensionful coupling constant $g_{4+n}$ must be rescaled to identify the four-dimensional coupling using the compactification scale, that is, $g=g_{4+n}f^{(\underline{0})}_E=g_{4+n}/(2\pi R)^{\frac{n}{2}}$.
\\

Continuing with the analogy between the KK and the Brout-Englert-Higgs mass-generating mechanisms, it should be recalled that in the SM any mass, emerged after SSB, is given as $\sigma v$, with $v$ the Fermi scale and $\sigma$ a dimensionless parameter, so that a single dimensionful scale is enough to define all the masses of the model. In the model there coexist light and heavy masses due to different values of the dimensionless parameters $\sigma$. This is an important feature of the BEHM. The KK mass-generating mechanism (KKM) also possesses this feature. In fact, if, as it has been done throughout the work, we assume in Eq.(\ref{MassA}) that all radii are equal,  $R_1=\cdots=R_n\equiv R$, we have
\begin{equation}
m_{(\underline{m})}=R^{-1}\sqrt{\underline{m}^2_1+\cdots +\underline{m}^2_n}\, ,
\end{equation}
which shows that the infinite KK mass spectrum is determined only by one dimensionful scale, namely, $R^{-1}$. As in the case of the BEHM, here some masses are heavier than others due to different values of dimensionless parameters, in this case, the diverse combinations of Fourier indices. So, for a fixed $R^{-1}$, we can have KK particles with increasing heavy masses by simply increasing the values of the Fourier indices. The lightest mass corresponds to the lowest configuration of Fourier indices. On the other hand, we can have a KK spectrum that is very heavy relative to the Fermi scale $v$ by letting $R\to 0$, which eventually would lead to the decoupling of these effects from low-energy observables, as it is established by the decoupling theorem~\cite{AC}.
\\

Previously, we have highlighted some similarities of the KKM with the BEHM. However, there is an essential difference between the two mechanisms that we would like to discuss now. For clarity, let us introduce a more compact notation. In terms of the Lie's basis $\{ T^a \}$, the gauge fields and gauge parameters of the $G$ group can be written as follows:
\begin{eqnarray}
\label{LBG}
A_\mu&=&A^a_\mu T^a \nonumber \\
&=&A^{\bar a}_\mu T^{\bar a} +A^{\hat a}_\mu T^{\hat a}\, ,
\end{eqnarray}
\begin{eqnarray}
\label{LBP}
\alpha&=&\alpha^a T^a \nonumber \\
&=&\alpha^{\bar a} T^{\bar a} +\alpha^{\hat a}T^{\hat a}\, .
\end{eqnarray}
On the other hand, in terms of the basis generated by the $P_{\bar \mu}$ and $\bar{P}^2$ observables, $\{|0 \rangle, \ |p^{(\underline{m})}\rangle \}$, the gauge fields and parameters of the ${\rm SU}(N,{\cal M}^d)$ group can be thought of as vectors of the space expanded by this basis as follows\footnote{This is in the same way that in non-relativistic quantum mechanics the state vector $|\alpha(t)\rangle$ is represented in a basis $\{|a\rangle \}$ generated by the observable $A$ as: $ |\alpha(t)\rangle=\sum_{a}c_a(t)|a\rangle$, with $x$ playing the role of time $t$.}:
\begin{subequations}
\begin{align}
\label{PBV}
|{\cal A}_\mu(x) \rangle &=A^{(\underline{0})}_\mu(x)|0\rangle +\sum_{(\underline{m})}A^{(\underline{m})}_\mu(x)|p^{(\underline{m})}\rangle\, , \\
\label{PBS}
|{\cal A}_{\bar \mu}(x) \rangle &=A^{(\underline{0})}_{\bar \mu}(x)|0\rangle +\sum_{(\underline{m})}A^{(\underline{m})}_{\bar \mu}(x)|p^{(\underline{m})}\rangle \, , \\
\label{PBP}
|\alpha (x) \rangle &=\alpha^{(\underline{0})}(x)|0\rangle +\sum_{(\underline{m})}\alpha^{(\underline{m})} (x)|p^{(\underline{m})}\rangle \, .
\end{align}
\end{subequations}
When the abstract expressions (\ref{PBV},\ref{PBS},\ref{PBP}) are represented in the coordinates basis $\{ |\bar x\rangle \}$ as $\langle\bar x|{\cal A}_\mu(x) \rangle={\cal A}_\mu (x,\bar x)$, etc., and the parity properties on their components are implemented, Eqs.(\ref{MFV22},\ref{MFS22},\ref{MGP2}) are straightforwardly recovered.
\\

In the context of the BEHM, assume that SSB $G\to H$ is implemented through the constant vector $\Phi_{0}$ that is left invariant by the $H$ subgroup. Then, the gauge fields associated with the broken generators $T^{\hat{a}}$ of $G$ acquire mass. The corresponding mass matrix emerges from the Higgs kinetic sector and is given by
\begin{eqnarray}
\label{MGEHM}
(D_\mu \Phi)^\dag (D^\mu \Phi)&=&g^2\left[\Phi^\dag_{0}\left(A_\mu A^\mu\right)\Phi_{0}\right]+\cdots\nonumber \\
&=&g^2\left[\Phi^\dag_{0}\left(\hat{A}_\mu \hat{A}^\mu\right)\Phi_{0}\right]+\cdots \nonumber \\
&=&g^2\left[\Phi^\dag_{0\, i}\left(\hat{A}_\mu \hat{A}^\mu\right)_{ij}\Phi_{0\, j}\right]+\cdots
\end{eqnarray}
On the other hand, in the case of the KK gauge fields, the corresponding mass matrix emerges from the KK kinetic sector as follows:
\begin{eqnarray}
\label{MGKKM}
\sum_{(\underline{m})}{\rm Tr}\left({\cal F}^{(\underline{m})}_{\mu \bar \nu}{\cal F}^{(\underline{m})\mu}\hspace{0.000001cm}_{\bar \nu}\right)&=&\sum_{(\underline{m})}p^{(\underline{m})}_{\bar \nu}p^{(\underline{m})}_{\bar \nu}\,{\rm Tr}\left(A^{(\underline{m})}_\mu A^{(\underline{m})\mu}\right)+\cdots \nonumber \\
&=&{\rm Tr}\left[\langle {\cal A}_{\mu}(x)|\bar{P}^2 |{\cal A}^\mu(x)\rangle \right]+\cdots \nonumber \\
&=&\sum_{(\underline{rs})}p^{(\underline{r})}_{\bar \nu}\,{\rm Tr}\left[\langle p^{(\underline{r})}|A^{(\underline{r})}_\mu A^{(\underline{s})\mu}|p^{(\underline{s})} \rangle  \right]p^{(\underline{s})}_{\bar \nu} +\cdots \, ,
\end{eqnarray}
where Tr denotes a trace on products of Lie generators, which we assume to be normalized as ${\rm Tr}(T^aT^b)=\delta^{ab}/2$.
\\

The structures of Eqs.~(\ref{MGEHM},\ref{MGKKM}) are quite suggestive. In the BEHM context, one says that associated with each broken generator $T^{\hat a}$ of the group $G$, there are a massive gauge boson $A^{\hat a}_\mu$, a pseudo-Goldstone boson $A^{\hat a}_{\rm G}$, and a gauge parameter $\alpha^{\hat a}$ defining a NSGT. On the other hand, in the case of the KKM, one says that associated with each eigenket $|p^{(\underline{m})}\big>$ of $P_{\bar \mu}$, there are a massive gauge boson $A^{(\underline{m})a}_\mu$, a pseudo-Goldstone boson $A^{(\underline{m})a}_{\rm G}$, and a gauge parameter $\alpha^{(\underline{m})a}$ defining a NSGT. Following with this analogy, one can say that associated with each unbroken generator $T^{\bar a}$ of $G$, there are a gauge field $A^{\bar a}_\mu$ and a gauge parameter $\alpha^{\bar a}$ defining a SGT; the set of generators $T^{\bar a}$ defining the Lie algebra of the subgroup $H$. On the other hand, in the KKM context, one says that associated with the base state $|0\big>$ of $P_{\bar \mu}$, there are a set of gauge fields $A^{(\underline{0})a}_\mu$ and a set of gauge parameters $\alpha^{(\underline{0})a}$ defining SGTs; both sets of gauge fields and gauge parameters defining the gauge group $SU(N,{\cal M}^4)$.
\\

Then, we have the following result: \textit{In the context of the KK mass-generating mechanism, we say that associated with each excited vector $\{ |p^{(\underline{m})}\big >\}$, of the basis generated by the Casimir invariant ${\bar P}^2$, there are a massive gauge boson, $A^{(\underline{m})a}_\mu$, a pseudo-Goldstone boson, $A^{(\underline{m})a}_{\rm G}$, and a gauge parameter, $\alpha^{(\underline{m})a}$; the standard gauge bosons $A^{(\underline{0})a}_\mu$ and standard gauge parameters $\alpha^{(\underline{0})a}$ being aligned along the base state $|0\big>$}.
\\

\section{Summary}
\label{C}
In this paper, we have presented an extra-dimensional extension of Yang-Mills theories in which $n$ flat compact extra dimensions are incorporated to define an effective theory. Our starting point has been a field theory that is valid at energies far above the compactification scale $R^{-1}$, which respects the extended  ${\rm ISO}(1,3+n)\otimes {\rm SU}(N,{\cal M}^d)$ symmetries. It is assumed that the sizes $R_i$ of the extra dimensions are so large, compared with the distance scales at which this theory is valid, that they can be practically considered as infinite.
\\

To describe the physical phenomena at energies of order of the compactification scale $R^{-1}$, we resort to the notion of hidden symmetry and to a mass-generating mechanism, in this case the Kaluza-Klein mass-generating mechanism or compactification. In order to hide the ${\rm ISO}(1,3+n)\otimes {\rm SU}(N,{\cal M}^d)$ symmetries into ${\rm ISO}(1,3)\otimes {\rm SU}(N,{\cal M}^4)$, two canonical transformations were implemented. First, a canonical map was implemented in order to accommodate ${\rm SO}(1,3+n)$ representations into ${\rm SO}(1,3)$ representations. This map allows one to hide the ${\rm SO}(1,3+n)$ symmetry into the ${\rm SO}(1,3)$ one. Next, a second nontrivial canonical map was implemented in order to hide any manifest dynamical role of the inhomogeneous ${\rm ISO}(n)$ group. Crucial to this map is the assumption of the existence of a set of orthonormal functions $\{ f^{(\underline{0})}, \, f^{(\underline{m})}(\bar x)\}$ defined on the compact manifold. The presence of the constant function $f^{(\underline{0})}$, which may be common to any compactification scheme, plays a central role in defining the connections and gauge parameters of the standard gauge group ${\rm SU}(N,{\cal M}^4)$. So, the components of ${\cal A}^a_\mu(x,\bar x)$ and $\alpha^a(x,\bar x)$ along $f^{(\underline{0})}$ can be identified, respectively, as the gauge fields and gauge parameters of the ${\rm SU}(N,{\cal M}^4)$ group, while their components along the $f^{(\underline{m})}(\bar x)$ directions emerge in the adjoint representation of this group. This map also allows one to identify the KK fields $A^{(\underline{m})a}_\mu(x)$ as genuine gauge fields because it turns out that there is a one-to-one relation with the gauge parameters $\alpha^{(\underline{m})a}(x)$. As far as the scalar fields  ${\cal A}^a_{\bar \mu}(x,\bar x)$ are concerned, their components, either along  $f^{(\underline{0})}$ or along $f^{(\underline{m})}(\bar x)$, transform in the adjoint representation of the ${\rm SU}(N,{\cal M}^4)$ group. The components of both ${\cal A}^a_\mu(x,\bar x)$ and ${\cal A}^a_{\bar \mu}(x,\bar x)$ fields along the $f^{(\underline{m})}(\bar x)$ directions can be endowed with mass, since they appear as matter fields from the ${\rm SU}(N,{\cal M}^4)$ group perspective. In general, this is enough to correctly identify the ${\rm ISO}(1,3)\times {\rm SU}(N,{\cal M}^4)$ covariant structure of the new basic fields. This means that in order to hide the extended symmetries into the standard ones, it is not necessary to specify the geometry of the compact manifold.
\\

The fact that the $f^{(\underline{0})}$ direction allows us to identify the connections of the standard gauge group, means that any class of field with component along this direction does not receive mass at the $R^{-1}$ scale. In contraposition, only those fields along the $f^{(\underline{m})}(\bar x)$ directions are endowed with mass by the Kaluza-Klein mechanism. In order to recover the known particle spectrum of the standard theory, the basis $\{ f^{(\underline{0})}, \, f^{(\underline{m})}(\bar x)\}$ was divided into two sub-bases: a basis of even functions, $\{ f^{(\underline{0})}, \, f^{(\underline{m})}_{\rm E}(\bar x)\}$, and a basis of odd functions, $\{ f^{(\underline{m})}_{\rm O}(\bar x)\}$. It was postulated that, with respect to the reflection $\bar{x}\to-\bar{x}$, fields with standard counterpart are necessarily even, while fields without standard counterpart are odd. The Kaluza-Klein mass-generating mechanism operates through the generators of the translations group ${\rm T}(n)$ of the inhomogeneous group ${\rm ISO}(n)$. Because of this, it is natural to define the set of orthonormal functions $\{ f^{(\underline{0})}, \, f^{(\underline{m})}_{\rm E}(\bar x), \, f^{(\underline{m})}_{\rm O}(\bar x)\}$ as the eigenfunctions of the Casimir invariant of ${\rm ISO}(n)$ associated with the translations. Parity was introduced by assuming a compact manifold made of $n$ copies of the orbifold $S^1/Z_2$. Dirichlet boundary conditions on the odd solutions were assumed, whereas Neumann boundary conditions were imposed on the even solutions. The eigenvalues of the Casimir invariant are the squared masses of the Kaluza-Klein particles. The eigenfunction associated with the zero eigenvalue corresponds to the constant $f^{(\underline{0})}$.
\\

We have shown that there is an interesting parallelism of the Kaluza-Klein mass-generating mechanism with the Brout-Englert-Higgs mechanism. Common to both mechanisms is the concept of hidden symmetry, which sets the stage to endow gauge fields with mass. In the former case, we perform a canonical map between two internal gauge groups $G  \xrightarrow{v} H$; while in the latter one, the corresponding map takes place between two spacetime groups, ${\rm ISO}(1,3+n)  \xrightarrow{R^{-1}} {\rm ISO}(1,3)$. However, as to the mechanism that allows us to generate the mass terms in the theory, we have found that there is an essential difference between the two approaches. In the case of the Brout-Englert-Higgs mechanism, SSB tells us that there are a massive gauge boson, a pseudo-Goldstone boson, and a gauge parameter associated with each broken generator of $G$; this set of gauge parameters define the NSGTs, which in turn allow us to define the unitary gauge. In the same context, associated with each unbroken generator of $G$, there are a gauge field and a gauge parameter; this kind of gauge parameters defines the gauge transformations associated with the subgroup $H$, which we have called SGT. In contrast, in the context of the Kaluza-Klein mechanism, we have shown that there are a massive gauge boson, a pseudo-Goldstone boson, and a gauge parameter associated with each vector $|p^{(\underline{m})}\neq 0\rangle$ of the basis generated by the Casimir invariant of the translations group ${\rm T}(n)\subset {\rm ISO}(n)$; the complete set of this type of gauge parameters defines the NSGTs, which, as in the standard case, serve to define unitary propagators for the KK gauge excitations. In this case, the gauge fields and gauge parameters of the standard group ${\rm SU}(N,{\cal M}^4)$ are aligned along the base state $|0\rangle$. On the other hand, both mechanisms share the property of generating a mass spectrum that is given as the product of a dimensionless constant by the dimensionful scale of the theory.

\appendix

\section{Multiple sums}
\label{S}
The symbol  $\sum_{(\underline{m})}$ summarizes a total of $ 2^{n}-1 $ different series and coincides with the notation $ \sum' $ used in Ref.~\cite{OPT3}. In fact,
\begin{align}
\sum_{(\underline{m})}T^{(\underline{m})}  &:=  \sum_{m_{1}=1}^{\infty}T^{(m_{1},0,\ldots,0)}+\sum_{m_{2}=1}^{\infty}T^{(0,m_{2},0,\ldots,0)}+\ldots+ \sum_{m_{n}=1}^{\infty}T^{(0,\ldots,m_n)} \nonumber \\
&+\sum_{m_{1},m_{2}=1}^{\infty}T^{(m_{1},m_{2},0,\ldots,0)}+\ldots+ \sum_{m_{n-1},m_{n}=1}^{\infty}T^{(0,\ldots,0,m_{n-1},m_n)} \nonumber \\
&\vdots \nonumber \\
& +\sum_{m_{1},\ldots, m_{n}=1}^{\infty}T^{(m_{1},\ldots, m_{n})}\ .\label{SD}
\end{align}
Since the positions of the Fourier indices in the spaces of $(\underline{m})$ are not relevant, but only the number of them that have been occupied, in practice one can use the following definition
\begin{equation}
\sum_{(\underline{m})}=
\sum^n_{l=1}\left(\begin{array}{ccc}
n \\
l
\end{array}\right)\sum^{\infty}_{(m_1,\cdots,m_l)=1}\, .
\end{equation}

\section{List of useful integrals and conventions}
\label{AE}

Recall that in Sec.~\ref{GF}, fields were defined on the base manifold $\mathcal{M}^d$, which is the product of the Minkowsky spacetime manifold in four dimensions, $\mathcal{M}^{4}$, and the compact manifold made of $n$ copies of the orbifold $S^{1}/Z_{2}$. These fields were Fourier expanded using their periodicity and parity. In order to make the notation manageable, the following constant and functions were defined:

\begin{align}
f^{(\underline{0})}_{E} & :=\frac{1}{\sqrt{(2\pi R_{1})\cdots(2\pi R_{n})}}\,  \\
f^{(\underline{m})}_{E}(\bar{p}\cdot \bar{x}) & := \sqrt{\frac{2}{(2\pi R_{1})\cdots(2\pi R_{n})}} \cos \left(\bar{p}\cdot \bar{x} \right)\, , \\
f^{(\underline{m})}_{O}(\bar{p}\cdot \bar{x}) & := \sqrt{\frac{2}{(2\pi R_{1})\cdots(2\pi R_{n})}} \sin \left(\bar{p}\cdot \bar{x} \right)\, ,
\end{align}
where $ \bar{p}\cdot \bar{x}=\frac{\underline{m}_1 \overline{x}_1}{R_1}+\ldots +\frac{\underline{m}_{n}\overline{x}_{n}}{R_{n}}$. To be precise, this scalar product should be written as $ \bar{p}^{(\underline{m})}\cdot \bar{x}$ in order to stress the fact that all possible combinations of $p_{\bar \mu}$ along the extra dimensions  must be considered. The underlines on Fourier indices in $\frac{\underline{m}_1 \overline{x}_1}{R_1}+\ldots +\frac{\underline{m}_{n}\overline{x}_{n}}{R_{n}}$ emphasizes this fact (see Ref.~\cite{OPT3})

The process of compactification involves integration on the extra dimensions $\overline{x}=\left(x_5,\cdots x_{4+n}\right)$, of certain combinations of products of the even and odd functions defined above. In this appendix we collect some useful integrals\footnote{Notice that the integral limits and the arguments of the even and odd functions used here are a reparametrization of those used in \cite{OPT3}. Also observe that throughout the paper we use, by simplicity,  $f^{(\underline{m})}(\bar x)$ instead of $f^{(\underline{m})}(\bar{p}\cdot \bar{x})$}.\\

\noindent Integral of one even/odd function $f^{(m)}$,
\begin{equation}
 \int_{0}^{2\pi R_n}\ldots\int_{0}^{2\pi R_1}\mathrm{d}^{n} \overline{x}\quad f^{(\underline{m})}_{E}(\bar{p}\cdot \bar{x})
=\int_{0}^{2\pi R_n}\ldots\int_{0}^{2\pi R_1}\mathrm{d}^{n}\overline{x} \quad f^{(\underline{m})}_{O}(\bar{p}\cdot \bar{x})=0\,.
 \end{equation}
\noindent The orthonormality of the set $\{f^{(\underline{m})}_{E}(\bar{p}\cdot \bar{x}),\,f^{(\underline{m})}_{O}(\bar{p}\cdot \bar{x})\}$,

\begin{eqnarray}
\int_{0}^{2\pi R_n}\ldots\int_{0}^{2\pi R_1}&\mathrm{d}^{n}\overline{x}& f^{(\underline{m})}_{E}(\bar{p}\cdot \bar{x})
 f^{(\underline{k})}_{O}(\bar{p}\cdot \bar{x}) =0\,, \\\\
\int_{0}^{2\pi R_n}\ldots\int_{0}^{2\pi R_1}&\mathrm{d}^{n}\overline{x}&f^{(\underline{m})}_{E}(\bar{p}\cdot \bar{x})f^{(\underline{k})}_{E}(\bar{p}\cdot \bar{x}) \\ \\
= \int_{0}^{2\pi R_n}\ldots\int_{0}^{2\pi R_1}&\mathrm{d}^{n}\overline{x}& f^{(\underline{m})}_{O}(\bar{p}\cdot \bar{x})f^{(\underline{k})}_{O}(\bar{p}\cdot \bar{x})
=\delta_{(\underline{m} \underline{k})}\, .
\end{eqnarray}

\noindent The integrals of combinations of the product of three even and/or odd functions with different Fourier modes,

\begin{equation}
\int_{0}^{2\pi R_n}\ldots\int_{0}^{2\pi R_1}\mathrm{d}^{n}\overline{x} f^{(\underline{m})}_{E}(\bar{p}\cdot \bar{x})f^{(\underline{k})}_{E}(\bar{p}\cdot \bar{x})f^{(\underline{r})}_{E}(\bar{p}\cdot \bar{x})
=f^{(\underline{0})}_E\,\,\Delta_{(\underline{m} \underline{k} \underline{r})}\, ,
\end{equation}

\begin{equation}
\int_{0}^{2\pi R_n}\ldots\int_{0}^{2\pi R_1}\mathrm{d}^{n}\overline{x} f^{(\underline{m})}_{O}(\bar{p}\cdot \bar{x})f^{(\underline{k})}_{O}(\bar{p}\cdot \bar{x})f^{(\underline{r})}_{E}(\bar{p}\cdot \bar{x})
=f^{(\underline{0})}_E\,\,\Delta'_{(\underline{m} \underline{k} \underline{r})}\, ,
\end{equation}

\begin{eqnarray}
\int_{0}^{2\pi R_n}\ldots\int_{0}^{2\pi R_1}&\mathrm{d}^{n}\overline{x}& f^{(\underline{m})}_{O}(\bar{p}\cdot \bar{x})f^{(\underline{k})}_{O}(\bar{p}\cdot \bar{x})f^{(\underline{r})}_{O}(\bar{p}\cdot \bar{x}) \\
=\int_{0}^{2\pi R_n}\ldots\int_{0}^{2\pi R_1}& \mathrm{d}^{n}\overline{x}& f^{(\underline{m})}_{E}(\bar{p}\cdot \bar{x})f^{(\underline{k})}_{E}(\bar{p}\cdot \bar{x})f^{(\underline{r})}_{O}(\bar{p}\cdot \bar{x})
=0\, ,
\end{eqnarray}
where
\begin{subequations}
\begin{align}
\Delta_{(\underline{rms})}&=\frac{1}{\sqrt{2}}\left(\delta_{\underline{m},\ \underline{r}+\underline{s}}+\delta_{\underline{r}, \ \underline{m}+\underline{s}}+\delta_{\underline{s}, \ \underline{r}+\underline{m}} \right)\, ,\\
\Delta'_{(\underline{rms})}&=\frac{1}{\sqrt{2}}\left(\delta_{\underline{m},\ \underline{r}+\underline{s}}+\delta_{\underline{r}, \ \underline{m}+\underline{s}}-\delta_{\underline{s}, \ \underline{r}+\underline{m}} \right)\, .
\end{align}
\end{subequations}

\section*{Acknowledgments}
We thank I. Garc\'\i a-Jim\' enez, A. Granados-Gonz\' alez, and G. I. N\' apoles-Ca\~nedo for their participation in early stages of the present work.
We acknowledge financial support from SECIHTI (Secretar\'\i a de Ciencia, Humanidades, Tecnolog\'\i a e Innovaci\' on) and SNII (Sistema Nacional de Investigadoras e Investigadores) (M\' exico).

\end{document}